\documentclass[11pt,a4paper]{article}

\usepackage{jcappub}
\usepackage{graphicx}

\usepackage{amssymb}
\usepackage{amsmath}
\usepackage{bm}
\usepackage{latexsym}
\usepackage{epsfig}
\usepackage{psfrag}
\usepackage[normalem]{ulem}
\usepackage{textcomp}
\usepackage{color}
\usepackage{pstricks}
\usepackage[utf8]{inputenc}

\makeatletter
\gdef\@fpheader{}
\makeatother

\def\nn{\nonumber} 
\def\pa{{\partial}}
\def\f{\frac}
\def\l{\left}
\def\r{\right}
\def\d{{\rm d}}
\def\Mpl{M_{_{\rm Pl}}}
\def\beq{\begin{equation}}
\def\eeq{\end{equation}} 
\def\beqa{\begin{eqnarray}}
\def\eeqa{\end{eqnarray}}

\def\cA{\mathcal A}
\def\cN{\mathcal N}

\def\Mp{M_{_{\rm Pl}}}

\def\cR{{\mathcal R}}
\def\cA{{\mathcal A}}
\def\cB{{\mathcal B}}
\def\cC{{\mathcal C}}
\def\cD{{\mathcal D}}
\def\cE{{\mathcal E}}
\def\cF{{\mathcal F}}
\def\cS{{\mathcal S}}
\def\cH{{\mathcal H}}
\def\cL{{\mathcal L}}

\newcommand{\g}{\gamma}

\newcommand{\viz}{\textit{viz.~}}
\newcommand{\ie}{\textit{i.e.~}}

\begin{document}
\title{Viable tensor-to-scalar ratio in a symmetric matter bounce}
\author[a]{Rathul Nath Raveendran,}
\affiliation[a]{The Institute of Mathematical Sciences, HBNI, CIT Campus, 
Chennai~600113, India}
\emailAdd{rathulnr@imsc.res.in}
\author[b]{Debika Chowdhury}
\affiliation[b]{Department of Physics, Indian Institute of Technology Madras, 
Chennai~600036, India}
\emailAdd{debika@physics.iitm.ac.in}
\author[b]{and L.~Sriramkumar}
\emailAdd{sriram@physics.iitm.ac.in}
\date{today}
\abstract
{Matter bounces refer to scenarios wherein the universe contracts at early 
times as in a matter dominated epoch until the scale factor reaches a minimum, 
after which it starts expanding. 
While such scenarios are known to lead to scale invariant spectra of primordial 
perturbations after the bounce, the challenge has been to construct {\it completely}\/ 
symmetric bounces that lead to a tensor-to-scalar ratio which is small enough to 
be consistent with the recent cosmological data. 
In this work, we construct a model involving two scalar fields (a canonical field 
and a non-canonical ghost field) to drive the symmetric matter bounce and study 
the evolution of the scalar perturbations in the model. 
We find that the model can be completely described in terms of
a single parameter, \viz\/ the ratio of the scale associated with 
the bounce to the value of the scale factor at the bounce.
We evolve the scalar perturbations numerically across the bounce and evaluate 
the scalar power spectra {\it after}\/ the bounce. 
We show that, while the scalar and tensor perturbation spectra are scale 
invariant over scales of cosmological interest, the tensor-to-scalar 
ratio proves to be much smaller than the current upper bound from the  
observations of the cosmic microwave background anisotropies by the
Planck mission.
We also support our numerical analysis with analytical arguments.}

\maketitle


\section{Introduction}

Bouncing models refer to scenarios wherein the universe undergoes a period 
of contraction until the scale factor attains a minimum value, whereupon it 
transits to the more standard phase of expansion. 
Such scenarios provide an alternative to the inflationary framework as they can 
also aid in overcoming the horizon problem associated with the conventional hot 
big bang model, in a fashion similar to inflation. 
Importantly, certain bouncing scenarios are also known to lead to nearly scale 
invariant spectra of primordial perturbations (see, for instance, the 
reviews~\cite{Novello:2008ra,Cai:2014bea,Battefeld:2014uga,Lilley:2015ksa,
Ijjas:2015hcc,Brandenberger:2016vhg}), as required by the cosmological data.
It is generally expected that quantum gravitational effects will have a
substantial influence on the dynamics of the very early universe, close 
to the big bang.
In this work, we shall consider {\it classical bounces},\/ which correspond to
situations wherein the background energy density remains sufficiently lower than 
the Planckian energy density, even as the universe evolves across the bounce. 
This enables us to carry out our analysis without having to take into account 
possible Planck scale effects, which may otherwise play a significant role 
near the bounce.

\par

Though there may be differences of opinion about the theoretical motivations for 
specific models, it has to be acknowledged that, as a broad paradigm, inflation 
has been a tremendous success (see, for example, the following 
reviews~\cite{Mukhanov:1990me,Martin:2003bt,Martin:2004um,Bassett:2005xm,
Sriramkumar:2009kg,Sriramkumar:2012mik,Baumann:2009ds,Linde:2014nna,Martin:2015dha}).
However, the remarkable effectiveness of the inflationary paradigm also seems to
be responsible for its major drawback. 
Despite the strong observational constraints that have emerged, we still seem far
from the desirable goal of arriving at a reasonably small subset of viable 
inflationary models (for a comprehensive list of single field models and their
performance against the cosmological data, see Refs.~\cite{Martin:2010hh,
Martin:2013tda,Martin:2013nzq,Martin:2014rqa}).
Moreover, it is not clear whether the paradigm can be falsified at all (in this
context, see Ref.~\cite{Gubitosi:2015pba})!
In sharp contrast, bouncing models have been plagued by various difficulties
and constructing a model that is free of pathologies, while being consistent with 
the observations, seems to pose considerable challenges. 
At this stage, we believe it is important that we highlight some of the generic 
issues.
Firstly, in a universe which is undergoing accelerated expansion, any classical 
perturbations that are originally present in the sub-Hubble regime will quickly 
decay.
But, such perturbations will rapidly grow during the contracting phase as one 
approaches the bounce.
This behavior raises the concern if a smooth and homogeneous background that 
is required as an initial condition at suitably early times is sufficiently 
probable.
It can also bring into question the validity of linear perturbation theory in the 
proximity of the bounce~\cite{Vitenti:2011yc,Pinto-Neto:2013zya,Battefeld:2014uga,
Brandenberger:2016vhg,Easson:2016klq}. 
However, it has been shown that, for a large class of bouncing scenarios, one can 
work in a specific, well-defined gauge wherein linear perturbation theory is valid 
near the bounce (in this context, see Refs.~\cite{Vitenti:2011yc,Pinto-Neto:2013zya}).
Secondly, small anisotropies are known to grow during the contracting phase, which 
may lead to the so-called Belinsky-Khalatnikov-Lifshitz instability~\cite{Belinsky:1970ew}.
While the above two issues can be alleviated to some extent in specific models such 
as the ekpyrotic scenario (see, for example, Refs.~\cite{Khoury:2001wf,Khoury:2001zk,
Lehners:2008vx,Levy:2015awa}), generically, they could be overcome only by careful 
fine tuning of the initial conditions (for a recent discussion on this point, see 
Ref.~\cite{Levy:2016xcl}).
Thirdly, certain gauge invariant quantities are bound to diverge in the vicinity 
of the bounce [when the Null Energy Condition (NEC) is initially violated and later 
restored], which may pose fundamental difficulties in evolving the perturbations 
across the bounce.
However, as we shall discuss in due course, this difficulty can be circumvented by 
working in a suitable gauge and evolving the perturbations in that particular gauge 
(in this context, see, for example, Ref.~\cite{Allen:2004vz}).
Fourthly, vector perturbations, if present, can grow rapidly in a contracting
universe~\cite{Battefeld:2004cd,Pinto-Neto:2013zya}.
But, this issue can be overcome if one assumes that there are no vector sources
at early times.
In spite of such issues, bouncing models have attracted a lot of attention in the
literature over the last two decades (for an intrinsically incomplete list of
efforts in this direction, see Refs.~\cite{Martin:2001ue,Khoury:2001wf,Khoury:2001zk,
Tsujikawa:2002qc,Peter:2002cn,Peter:2003rg, Martin:2003sf,Levy:2016xcl,Allen:2004vz,
Bozza:2005wn,Cai:2007qw,Finelli:2007tr,Cai:2007zv,Lehners:2008vx,Battefeld:2004cd,
Levy:2015awa,Falciano:2008gt,Cai:2008qw,Lin:2010pf,Easson:2011zy,Cai:2011zx,Qiu:2011cy,
Cai:2012va,Cai:2013kja,Cai:2013vm,Cai:2014xxa,Odintsov:2014gea,Gao:2014eaa,Quintin:2015rta, 
Nojiri:2016ygo,Quintin:2016qro,Ijjas:2016tpn,Fertig:2016czu}).
These efforts suggest that bouncing scenarios can be regarded as the most popular 
alternative to the inflationary paradigm.
In this work, we shall consider a specific model leading to a completely symmetric 
matter bounce and investigate, both numerically and analytically, the evolution of 
scalar perturbations in this scenario.

\par

A matter bounce corresponds to a certain class of bouncing scenarios wherein,
during the early stages of the contracting phase, the scale factor behaves as 
in a matter dominated universe. 
Such models are known to be `dual' to de Sitter inflation, and hence are expected 
to lead to scale invariant spectra of primordial perturbations~\cite{Wands:1998yp,
Wands:2008tv}.
Before we go on to discuss about the specific model that we shall consider, let us 
make a few summarizing remarks regarding the existing matter bounce models.
One of the primary problems concerning symmetric matter bounce scenarios seems to 
be the fact that in many of these models~\cite{Allen:2004vz,Battefeld:2014uga,
Cai:2014bea,Cai:2014xxa, Cai:2008qw}, the tensor-to-scalar ratio $r$ turns out to 
be far in excess of the current upper bound of $r \lesssim 0.07$ from the Planck
mission~\cite{Ade:2015lrj}.
One possible way of circumventing this difficulty seems to be to 
model the regular component as a perfect fluid.
In particular, a suitably small speed of sound for the scalar perturbations ensures 
that the tensor-to-scalar ratio $r$ is small enough to be consistent with the 
data~\cite{Falciano:2007yf,Peter:2008qz}.
Due to the small speed of sound, the scalar perturbations leave the Hubble radius at 
earlier times (when compared with the tensor perturbations) providing them with more 
time for their amplitude to grow as the bounce is approached.
It has been also been shown that, by making a judicious choice of the initial 
conditions, a small tensor-to-scalar ratio can be obtained in asymmetric 
bounces~\cite{Allen:2004vz}.
Within the context of Einsteinian gravity, it is well known that the NEC has to be 
violated in order to obtain a bounce.
Moreover, since the Hubble parameter changes sign at the bounce, the total background 
energy density vanishes at the bounce.
The simplest way to drive such a background would be to introduce a ghost field 
which carries a negative energy density (see, for instance, Refs.~\cite{Allen:2004vz,
Cai:2007qw,Cai:2007zv}).
However, there are certain issues associated with ghost fields, mostly pertaining 
to the absence of a stable quantum vacuum~\cite{Cline:2003gs}.
The so-called ghost-condensate mechanism has been introduced as an improvement upon 
the typical ghost fields because the perturbative ghost instability can be avoided in 
this situation (see, for example, Refs.~\cite{Lin:2010pf, Fertig:2016czu}).
Nevertheless, it has been shown that it is impossible to embed the ghost condensate 
Lagrangian into an ultraviolet complete theory~\cite{ArkaniHamed:2003uy}.
In the matter bounce curvaton scenario~\cite{Cai:2011zx}, which also contains a ghost 
field in addition to a much lighter second field, the tensor-to-scalar ratio has been 
shown to be suppressed by `kinetic amplification'.
Another alternative would be to use the Galileon Lagrangian~\cite{Easson:2011zy,
Qiu:2011cy,Ijjas:2016tpn}, wherein the gradient instability, which may otherwise 
lead to an exponential growth of the comoving curvature perturbation, can be avoided. 
In certain single field models which lead to a non-singular bounce, it has been 
argued that the scalar perturbations are amplified more during the bounce relative 
to the tensor perturbations, which may lead to a viable value of~$r$~\cite{Cai:2012va,
Quintin:2015rta}.

\par

Therefore, the challenge seems to be to construct a {\it completely}\/ symmetric matter 
bounce scenario wherein the tensor-to-scalar ratio is small enough to be in agreement 
with the observations.
In an earlier work, we had studied the behavior of the tensor perturbations in a matter 
bounce scenario described by a specific form of the scale factor and had gone on to
evaluate the tensor power spectrum and bispectrum in the model~\cite{Chowdhury:2015cma}.
The most dominant of the primordial perturbations are, of course, the scalar perturbations.
While the tensor perturbations are completely determined by the behavior of the scale factor, 
as is well known, the evolution of the scalar perturbations strongly depends on the nature 
of the source driving the background.
In this work, assuming Einsteinian gravity, we shall construct a model that leads to 
the specific form of the scale factor for which we had previously obtained a scale 
invariant spectrum of tensor perturbations.
As we shall see, the scale factor of our interest can be driven with the aid of two 
scalar fields, one of which is a canonical field described by a potential, whereas 
the other is a purely kinetic ghost field with a negative energy density.
We shall show that it is possible to construct exact analytical solutions for the 
background dynamics of our model.
Utilizing the analytical solutions for the background and, working in a specific gauge,
we shall numerically evolve the perturbations across the bounce and evaluate the power 
spectrum of the scalar perturbations after the bounce. 
Interestingly, we find that the amplitude of the scale invariant scalar and tensor 
perturbation spectra (over cosmological scales) are dependent on only one parameter, 
\viz\/ the ratio of the scale associated with bounce to the value of scale factor at 
the bounce.
Further, we shall illustrate that the tensor-to-scalar ratio is completely independent 
of even this parameter, and it is in agreement with the constraints from Planck.
Lastly, we should mention that, we shall also present analytical arguments to support 
our numerical results.

\par

This paper is organized as follows. 
In the following section, we shall quickly introduce the scale factor characterizing
the bouncing scenario of our interest and stress a few basic points.  
In Sec.~\ref{sec:tp}, to illustrate some aspects, we shall revisit the behavior 
of the tensor perturbations and the evaluation of the tensor power spectrum in the 
scenario.
In Sec.~\ref{sec:model}, we shall construct the source for the bouncing scenario of 
our interest using two scalar fields.
In Sec.~\ref{sec:eom-sp}, we shall first arrive at the equations of motion 
governing the scalar perturbations in a generic gauge. 
Thereafter, we shall obtain the reduced equations in a specific gauge wherein the 
perturbations behave well in the vicinity of the bounce.  
In Sec.~\ref{sec:n}, we shall evolve the scalar perturbations numerically 
across the bounce.
In Sec.~\ref{sec:a}, we shall construct analytical solutions to the equations
governing the perturbations under certain approximations and we shall show 
that the analytical arguments support our numerical results.
In Sec.~\ref{sec:sps-r}, we shall evaluate the scalar power spectrum and the tensor-to-scalar
ratio after the bounce, both numerically as well as analytically, and illustrate that 
the resulting spectra are broadly in agreement with the constraints from the Planck data.
We shall conclude in Sec.~\ref{sec:so} with a summary and a brief outlook.

\par

We shall work with natural units such that $\hbar=c=1$, and set the Planck mass 
to be $\Mpl=\l(8\,\pi\, G\r)^{-1/2}$. 
We shall adopt the metric signature of $\l(-, +, +, +\r)$. 
Note that, while Greek indices shall denote the spacetime coordinates, the Latin 
indices shall represent the spatial coordinates, except for $k$ which shall be 
reserved for denoting the wavenumber. 
Moreover, an overdot and an overprime shall denote differentiation with respect 
to the cosmic and the conformal time coordinates, respectively. 
We shall also work with a new time variable that we have introduced in an earlier
work on bouncing scenarios, \viz e-$\cN$-folds, which we denote as 
$\cN$~\cite{Sriramkumar:2015yza,Chowdhury:2015cma}.


\section{The scale factor describing the matter bounce}\label{sec:sf}

We shall consider the background to be the spatially flat, 
Friedmann-Lema\^itre-Robertson-Walker (FLRW) metric that is described by 
the line element
\begin{equation}
\d s^2 = -\d t^2 + a^2(t)\,\delta_{ij}\, \d x^i\,\d x^j
=a^2(\eta)\, \l(-\d\eta^2+\delta_{ij}\, \d x^i\,\d x^j\r),\label{eq:flrw-le}
\end{equation}
where $a(t)$ is the scale factor and $\eta=\int \d t/a(t)$ denotes the 
conformal time coordinate.
We shall assume that the scale factor describing the bounce is given in 
terms of the conformal time as follows:
\begin{equation}
a(\eta) = a_0\l(1 + \eta^2/\eta_0^2\r) = a_0\l(1 + k_0^2\,\eta^2\r),
\label{eq:sf}
\end{equation}
where $a_0$ is the value of the scale factor at the bounce (\ie at 
$\eta=0$) and $k_0=1/\eta_0$ is the scale associated with the bounce. 
At very early times, \viz when $\eta \ll -\eta_0$, the scale factor behaves 
as $a \propto \eta^2$, which is the behavior in a matter dominated universe. 
It is for this reason that the above scale factor corresponds to a matter bounce
scenario.
In the absence of any other scale in the problem, it seems natural to assume
that the quantity $k_0$ is related to the Planck scale. 
As we shall see later, the source driving the scale factor
of our interest as well as the results we obtain depend only on the ratio 
$k_0/a_0$.
Specifically, it is the dimensionless ratio $k_0/(a_0\,\Mpl)$ that shall 
determine the amplitude of the power spectra.
We find that the scales of cosmological interest are about $20$-$30$ orders
of magnitude smaller than the wavenumber $k_0$ (in this context, see
Ref.~\cite{Chowdhury:2015cma}).

\par

Let us now highlight a few points concerning the above scale factor and the 
nature of the sources that are expected to drive the bounce.
To begin with, the scale factor is completely symmetric about the bounce.
Also, since the Hubble parameter $H=a'/a^2$ vanishes at the bounce, so does
the total energy density, \ie $\rho=3\,H^2\,\Mpl^2$, of the sources driving 
the scale factor.
It is straightforward to show that the energy density $\rho$ too is symmetric 
about the bounce.
The energy density initially increases on either side as one moves away from 
the bounce, reaches the maximum value $\rho_{\rm max}=3^4\, \Mpl^2\, k_0^2/(4^3\, 
a_0^2)$ at $\eta =\pm \eta_\ast$, where $\eta_\ast=\eta_0/\sqrt{3}$, and decreases 
thereafter.
Note that $\rho_{\rm max}$ depends only on the combination $k_0/a_0$.
The fact that $k_0/a_0$ is the only parameter in the problem will become more 
evident when we attempt to model the sources that drive the bounce.
If we assume $k_0/(a_0\,\Mpl)$ to be, say, of the order of $10^{-5}$ or so, then, 
clearly, the energy density $\rho$ will always remains much smaller than the 
Planckian density.
It is for this reason that we are able to treat the bounce as completely
classical.
Interestingly, in the domain $-\eta_\ast<\eta<\eta_\ast$, wherein the energy 
decreases as one approaches bounce, one finds that ${\dot H}>0$.
Since ${\dot H}=-(\rho+p)/(2\,\Mpl^2)$, where $p$ is the total pressure, 
$(\rho+p)<0$ during this period. 
In other words, the NEC is violated over this domain.
It should be clarified that, while $\eta_\ast\simeq 1/k_0$, the 
duration of the bounce in terms of the in terms of cosmic time is actually of the
order of $a_0/k_0$.

\par

It can be easily shown that the above scale factor can be driven by two fluids,
one which is ordinary, pressureless matter and another which behaves exactly as 
radiation, albeit with a negative energy density~\cite{Finelli:2007tr,Chowdhury:2016aet}.
In fact, it is this negative energy density (and the associated negative pressure) 
that leads to the violation of the NEC near the bounce and also ensures that
the total energy density of the two fluids vanishes at the bounce.
In due course, we shall model these two fluids in terms of scalar fields.


\section{The evolution of tensor perturbations and the tensor power 
spectrum}\label{sec:tp}

In this section, we shall revisit the evolution of the tensor perturbations
and the evaluation of the corresponding power spectrum in the matter bounce
scenario of our interest, which we have discussed in an earlier 
work~\cite{Chowdhury:2015cma}.
We shall study the evolution of the perturbations analytically as well as 
numerically.
This exercise permits us to introduce the concept of e-$\cN$-folds and also
highlight a few points concerning the evolution of perturbations in bouncing 
scenarios.
Later, we shall adopt similar methods to obtain analytical solutions for the 
scalar perturbations.
As we have emphasized earlier, the tensor perturbations are simpler to study 
because of the fact that the equation governing their evolution depends only 
on the scale factor.


\subsection{Analytical evaluation of the tensor perturbations}

Let us first discuss the analytical evaluation of the tensor modes and the 
tensor power spectrum.

\par

When the tensor perturbations characterized by $\g_{ij}$ are taken into account, 
the spatially flat FLRW metric can be expressed as (see, for instance,
Refs.~\cite{Mukhanov:1990me,Martin:2004um,Bassett:2005xm,Sriramkumar:2009kg,
Sriramkumar:2012mik})
\begin{equation}
\d s^2 =a^{2}(\eta)\, \l\{-\d \eta^2 
+ \l[\delta_{ij} + \gamma_{ij}(\eta, {\bm x})\r]
\d x^i\, \d x^j\r\}.\label{eq:flrw-le-wtp}
\end{equation}
The Fourier modes $h_k$ corresponding to the tensor perturbations are governed
by the differential equation
\begin{equation}
h_k''+2\,\f{a'}{a}\,h_k'+k^2\,h_k=0,\label{eq:eom-hk-eta}
\end{equation}
where, recall that, the overprimes denote differentiation with respect to the 
conformal time~$\eta$.
It proves to be convenient to introduce the so-called Mukhanov-Sasaki variable 
$u_k$ defined through the relation:~$h_k=\sqrt{2}\,u_k/(\Mpl\,a)$. 
The variable $u_k$ satisfies the differential equation 
\begin{equation}
u_k''+\l(k^2-\f{a''}{a}\r)\,u_k=0.\label{eq:eom-uk-eta}
\end{equation}
In the context of inflation, one imposes the standard Bunch-Davies initial 
condition on the modes when they are well inside the Hubble radius.
As we shall soon discuss, in bouncing scenarios, such a condition can
be imposed at sufficiently early times during the contracting phase.
The tensor power spectrum, evaluated at a suitably late time, say, $\eta_{\rm e}$,
is defined as \begin{equation}
{\cal P}_{_{\rm T}}(k) = 4\,\frac{k^3}{2\,\pi^2}\,\vert h_k(\eta_{\rm e})\vert^2.
\label{eq:tps-d}
\end{equation}
As is common knowledge, in the inflationary scenario, the power spectra are 
evaluated on super Hubble scales.
In bouncing models, the spectra are typically evaluated some time after
the bounce, when the universe is expected to make a transition to the
conventional radiation dominated epoch.

\par

From the expression~(\ref{eq:sf}) for the scale factor, we obtain that
\begin{equation}
\f{a''}{a}=\f{2\,k_0^2}{1+k_0^2\,\eta^2}.
\end{equation}
Clearly, the quantity $a''/a$ exhibits a maximum at the bounce, with the 
maximum value being of the order of $k_0^2$, and it vanishes as $\eta\to\pm 
\infty$. 
For modes of cosmological interest such that $k\ll k_0$, we find that 
$k^2\gg a''/a$ as $\eta\to-\infty$, \ie at very early times.
This behavior permits us to impose the standard Bunch-Davies initial 
condition on the modes $u_k$ at early times.

\par

As we mentioned, we shall be interested in evaluating the tensor power spectrum 
after the bounce.
Let us assume that, after the bounce, the universe transits to the radiation 
domination epoch at, say, $\eta=\beta\,\eta_0$, where we shall set $\beta
\simeq 10^2$.
We should hasten to clarify that, while this value of $\beta$ is somewhat arbitrary, 
we find that the final results do not strongly depend on the choice of $\beta$.
In order to study the evolution of the tensor modes analytically, let us divide
the period $-\infty< \eta \le \beta\,\eta_0$ into two domains.
The first domain is determined by the condition $-\infty<\eta \le -\alpha\,\eta_0$, 
where $\alpha$ is a very large number, which we shall set to be $10^5$. 
In other words, this domain corresponds to very early times during the contracting
phase before the bounce.
The second domain $-\alpha\,\eta_0\le \eta \le\beta\,\eta_0$ evidently involves 
periods prior to as well as immediately after the bounce.
We find that, under suitable approximations, we can evaluate the tensor modes
analytically in both of these domains.

\par

In the first domain (\ie during $-\infty<\eta \le -\alpha\,\eta_0$), the scale 
factor~(\ref{eq:sf}) reduces to
\begin{equation}
a(\eta)\simeq a_0\, k_0^2\, \eta^2,\label{eq:sf-d1}
\end{equation}
so that we have  $a''/a\simeq 2/\eta^2$, which is exactly the behavior in de
Sitter inflation. 
The Bunch-Davies initial condition that are to be imposed on the mode $u_k$ at 
early times when $k^2\gg 2/\eta^2$ is given by~\cite{Bunch:1978yq}
\begin{equation}
u_k=\f{1}{\sqrt{2\,k}}\, {\rm e}^{-i\,k\,\eta}.\label{eq:uk-ic}
\end{equation}
The modes $h_k$ that satisfy this initial condition in the first domain can be easily
determined to be~\cite{Starobinsky:1979ty,Wands:1998yp,Finelli:2001sr,Chowdhury:2015cma}
\begin{equation}
h_k\simeq\f{\sqrt{2}}{\Mpl}\,\f{1}{\sqrt{2\,k}}\,\f{1}{a_0\,k_0^2\,\eta^2}\,
\l(1-\f{i}{k\,\eta}\r)\,{\rm e}^{-i\,k\,\eta}.\label{eq:hk-d1}
\end{equation}

\par

Let us now consider the behavior of the modes in the second domain, 
\ie $-\alpha\,\eta_0\le\eta\le\beta\,\eta_0$.
In this domain, for scales of cosmological interest, which correspond to 
$k\ll k_0$, the equation governing the tensor mode $h_k$ simplifies to  
\begin{equation}
h_k''+\frac{2\,a'}{a}\,h_k'\simeq 0.\label{eq:eom-hk-d2}
\end{equation}
We should clarify that, since we are working in the domain wherein $\eta\ge
-\alpha\,\eta_0$, this equation is actually valid for wavenumbers such 
that $k\ll k_0/\alpha$.
The above equation can be integrated to yield
\begin{equation}
h_k(\eta)
\simeq h_k(\eta_1)+h_k'(\eta_1)\,a^2(\eta_1)\,
\int_{\eta_1}^{\eta} \frac{{\rm d}{\tilde \eta}}{a^2({\tilde \eta})},
\label{eq:hk-shs}
\end{equation}
where $\eta_1$ is a suitably chosen time, and we have set the constants 
of integration to be $h_k(\eta_1)$ and $h_k'(\eta_1)$.
Upon choosing $\eta_1=-\alpha\,\eta_0$ and using the form~(\ref{eq:sf})
of the scale factor, we find that, in the second domain, the tensor mode 
can be expressed as
\begin{equation}
h_k= {\cal A}_k+{\cal B}_k\,f(k_0\,\eta),\label{eq:hk-d2}
\end{equation}
where 
\begin{equation}
f(k_0\,\eta)=\f{k_0\,\eta}{1+k_0^2\,\eta^2}+\tan^{-1}\l(k_0\,\eta\r).
\label{eq:f}
\end{equation}
The quantities $\cA_k$ and $\cB_k$ can be determined from the solution~(\ref{eq:hk-d1})
in the first domain and are given by
\begin{subequations}\label{eq:Ak-Bk}
\begin{eqnarray}
{\cal A}_k &=& \f{\sqrt{2}}{\Mpl}\,\f{1}{\sqrt{2\,k}}\,\frac{1}{a_0\,\alpha^2}\,
\l(1+\frac{i\,k_0}{\alpha\,k}\right)\,{\rm e}^{i\,\alpha\,k/k_0}
+ {\cal B}_k\, f(\alpha),\\
{\cal B}_k &=& \f{\sqrt{2}}{\Mpl}\,\f{1}{\sqrt{2\,k}}\,\frac{1}{2\,a_0\,\alpha^2}\,
\l(1+\alpha^2\r)^2\,\l(\f{3\,i\,k_0}{\alpha^2\,k}+\frac{3}{\alpha}
- \frac{i\,k}{k_0}\r)\,{\rm e}^{i\,\alpha\,k/k_0}.
\end{eqnarray}
\end{subequations}
It is interesting to note here that, after the bounce, the first term in 
$f(k_0\,\eta)$ decays while the second term exhibits a mild growth.


\subsection{E-$\cN$-folds and the numerical evaluation of the tensor modes}

To understand the accuracy of the approximations involved, we can compare the 
above analytical results for the evolution of the tensor modes with the 
corresponding numerical results. 
Since the scale factor is specified, it is essentially a matter of numerically 
integrating the differential equation~(\ref{eq:eom-hk-eta}) governing the 
tensor perturbations with known coefficients.  
However, the conformal time coordinate does not prove to be an efficient time 
variable for numerical integration, in particular, when a large range in the 
scale factor needs to be covered.
In the context of inflation, one works with e-folds $N$ as the independent time
variable, with the scale factor being given by $a(N)\propto {\rm e}^N$.
But, the function ${\rm e}^N$ is monotonically increasing function, whereas, in 
a bounce, the scale factor decreases at first before beginning to increase.

\par

In order to describe the completely symmetric bouncing universe of our interest, 
we shall introduce a new time variable $\cN$, in terms of which the scale factor 
is given by~\cite{Sriramkumar:2015yza,Chowdhury:2015cma}
\begin{equation}\label{eq:scale-factor-N}
a(\cN) = a_0\,{\rm e}^{\cN^2/2}.
\end{equation}
We shall refer to the variable $\cN$ as e-$\cN$-fold, and we shall perform the 
numerical integration using this variable.
We shall assume that $\cN$ is zero at the bounce, with negative values 
representing the phase prior to the bounce and positive values after.

\par

\begin{figure}[!t]
\begin{center}
\includegraphics[width=12.00cm]{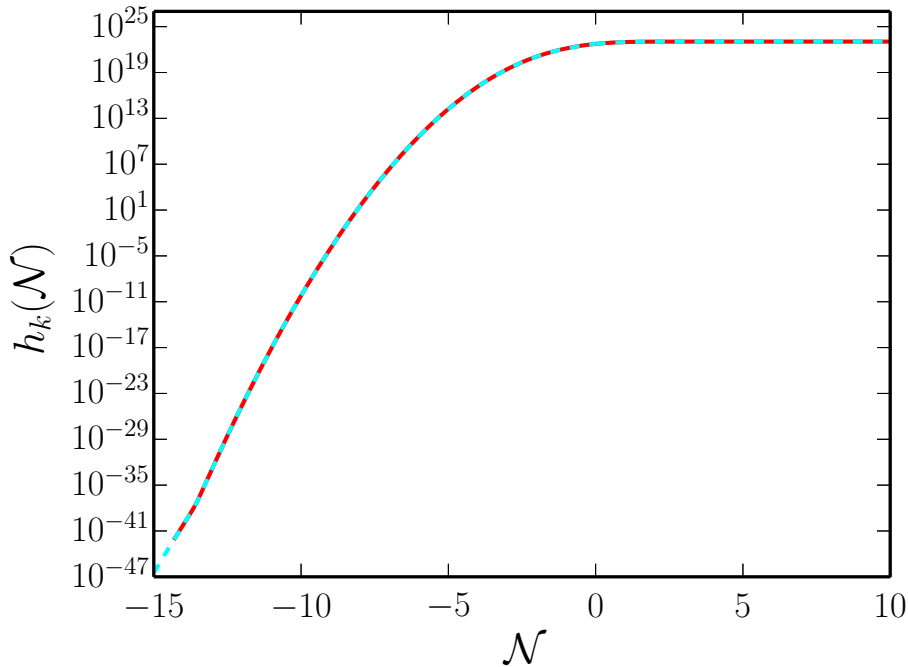}
\end{center}
\caption{The numerical (in red) and the analytical (in cyan) results for the 
amplitude of the tensor mode $h_k$ corresponding to the wavenumber $k/k_0 = 
10^{-20}$ has been plotted as a function of e-$\cN$-fold.
We have set $k_0/(a_0\,\Mpl)= 3.3\,\times\,10^{-8}$ and, for plotting the 
analytical results, we have also chosen $\alpha=10^5$.
Note that, to arrive at the plots we have chosen $k_0=\Mpl$ and $a_0=3.0\times10^7$, 
which is consistent with the abovementioned value of $k_0/(a_0\,\Mpl)$.
We have plotted the results from the initial e-$\cN$-fold $\cN_i$ [when $k^2
=10^4\, (a''/a)$] corresponding to the mode.
Clearly, the match between the analytical and numerical results is very good.
This indicates that the approximation for determining the modes analytically
works quite well.}\label{fig:hk}
\end{figure}

In terms of e-$\cN$-folds, the differential equation~(\ref{eq:eom-hk-eta}) 
governing the evolution of the tensor modes can be expressed as
\begin{equation}
\f{\d^2 h_k}{\d\cN^2}
+\l(3\,\cN+\f{1}{H}\,\f{\d H}{\d\cN}-\f{1}{\cN}\r)\,\f{\d h_k}{\d\cN}
+\l(\f{k\,\cN}{a\,H}\r)^2\,h_k=0,\label{eq:eom-hk-cN}
\end{equation}
where $H$ is the Hubble parameter.
In order to determine the coefficients of the above equation, we need to express 
the Hubble parameter in terms of e-$\cN$-folds.
Upon using the expression for the scale factor~(\ref{eq:sf}), we obtain that
\begin{equation}
\eta(\cN)= \pm\, k_0^{-1}\, \l({\rm e}^{\cN^2/2}-1\r)^{1/2}.
\end{equation}
It is important to note that, since the Hubble parameter is negative during the 
contracting phase and positive during the expanding regime, we shall have to 
choose the root of $\eta(\cN)$ accordingly during each phase.
We numerically integrate the differential equation~(\ref{eq:eom-hk-cN}) using 
a fifth order Runge-Kutta algorithm.
We impose the initial conditions at a sufficiently early time, say,  $\cN_i$, 
when $k^2=10^4\, (a^{\prime\prime}/a)$.
Evidently, the standard Bunch-Davies initial condition on $u_k$ 
[cf.~Eq.~(\ref{eq:uk-ic})] can be converted to initial conditions on the mode 
$h_k$ and its derivative with respect to the e-$\cN$-fold~\cite{Chowdhury:2015cma}.
The tensor mode $h_k$ evaluated numerically has been plotted in Fig.~\ref{fig:hk}
for a specific wavenumber.
In the same figure, we have also plotted the analytical result we have obtained
for the tensor mode.
It is clear from the figure that the analytical results match the exact numerical 
results exceedingly well, which illustrates the extent of accuracy of the 
analytical approximations.


\subsection{Tensor power spectrum}

\begin{figure}[!tbp] 
\begin{center}
\includegraphics[width=12.00cm]{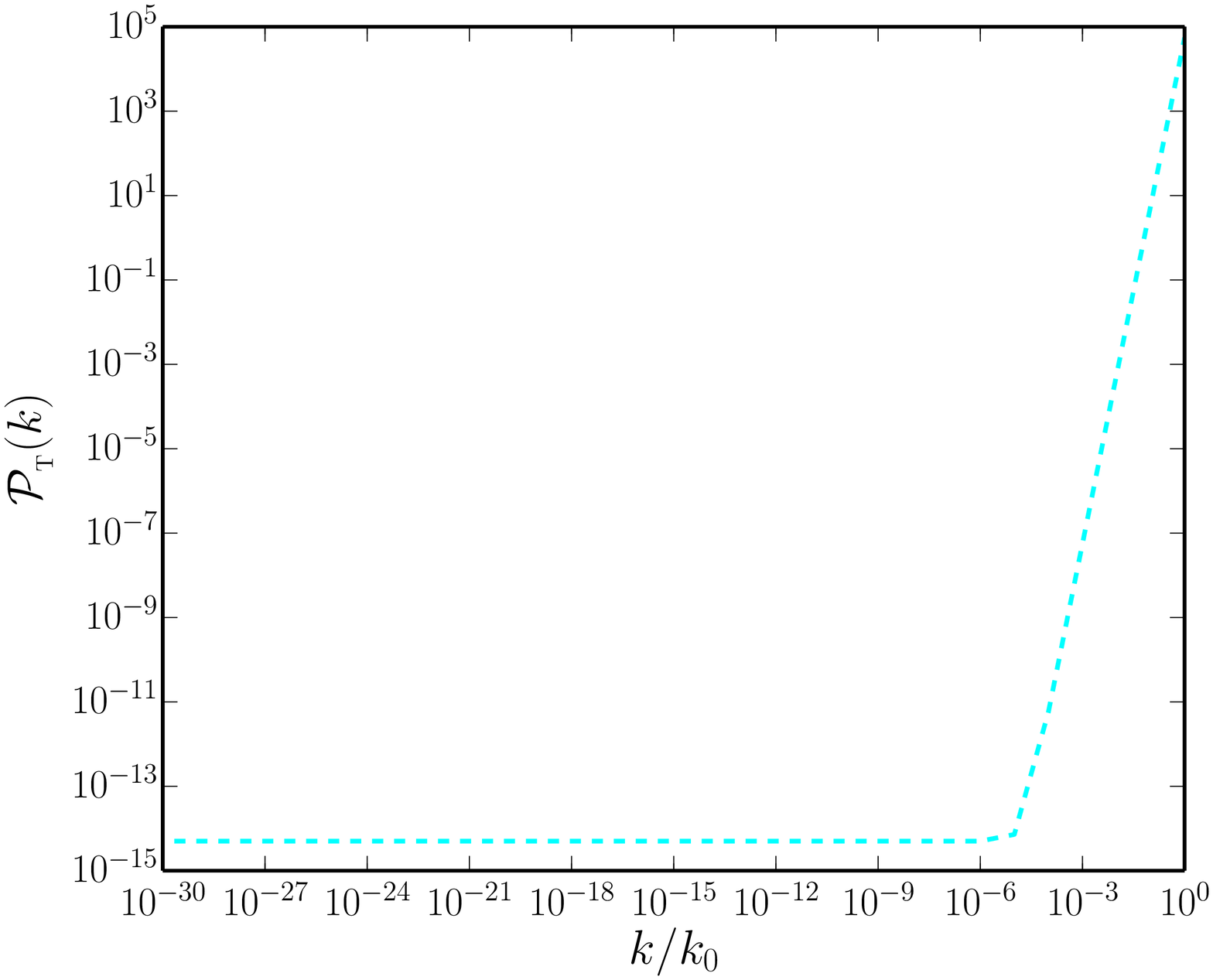}
\end{center}
\vskip -20pt
\caption{The tensor power spectrum ${\cal P}_{_{\rm T}}(k)$, evaluated 
analytically, has been plotted as a function of $k/k_0$ for a wide range 
of wavenumbers. 
In plotting this figure, we have chosen the same values for $k_0/a_0$ and 
$\alpha$ as in the previous figure, and have set $\beta=10^2$.
We should stress that the approximations we have worked with are valid
only for wavenumbers such that $k\ll k_0/\alpha$.
It is clear from the figure that the power spectrum is scale invariant 
over these wavenumbers.
Note that, for the values of the parameters mentioned above, at small 
enough wavenumbers, the tensor power spectrum has the scale invariant 
amplitude of ${\cal P}_{_{\rm T}}(k)\simeq 5\times 10^{-15}$.}
\label{fig:tps-a}
\end{figure}

The tensor power spectrum can now be evaluated using the solutions for the 
modes that we have obtained.
Upon substituting the modes~(\ref{eq:hk-d2}) in the expression~(\ref{eq:tps-d}), 
we find that the tensor power spectrum after the bounce, evaluated at $\eta=\beta
\,\eta_0$, can be written as
\begin{equation}
{\cal P}_{_{\rm T}}(k) 
= 4\,\frac{k^3}{2\,\pi^2}\,\vert {\cal A}_k+{\cal B}_k\, f(\beta)\vert^2,
\label{eq:tps-sia}
\end{equation}
with $\cA_k$ and $\cB_k$ given by Eqs.~(\ref{eq:Ak-Bk}), and $f$ by Eq.~(\ref{eq:f}).
As we had pointed out, our approximations are valid only for modes such that 
$k\ll k_0/\alpha$.
Also, for reasons discussed earlier, we need to choose $\beta$ to be reasonably 
large.
We have plotted the resulting tensor power spectrum in Fig.~\ref{fig:tps-a} for
$k_0/(a_0\,\Mpl)= 3.3\times10^{-8}$, $\alpha=10^5$ and $\beta=10^2$.
Clearly, the spectrum is scale invariant for wavenumbers such that 
$k\ll k_0/\alpha$.
It is straightforward to determine the scale invariant amplitude of the power
spectrum to be~\cite{Starobinsky:1979ty,Wands:1998yp,Finelli:2001sr}
\begin{equation}
{\cal P}_{_{\rm T}}(k)\simeq \f{9\,k_0^2}{2\,\Mpl^2\,a_0^2}.\label{eq:ta}
\end{equation}
Note that this tensor power spectrum depends only on the parameter $k_0/a_0$.
As we shall illustrate later (see Fig.~\ref{fig:ps-n}), the numerical results 
for the power spectrum (evaluated at $\eta=\beta\, \eta_0$) matches this scale 
invariant amplitude quite well.


\section{Modeling the bounce with scalar fields}\label{sec:model}

Our aim now is to construct sources involving scalar fields to drive the 
scale factor~(\ref{eq:sf}).
We had mentioned earlier that the scale factor can be achieved with the aid 
of two fluids, one of which is pressureless matter and another which behaves
as radiation, but with a negative energy density. 
It is well known that non-canonical scalar fields with a purely kinetic term
can act as perfect fluids~\cite{Garriga:1999vw,Akhoury:2008nn,Unnikrishnan:2010ag,
Pujolas:2011he,PhysRevD.81.107301}.
However, purely kinetic scalar fields cannot mimic pressureless matter, as a 
potential term is required to ensure that the pressure always remains zero.
We shall model pressureless matter by a canonical scalar field with a potential, 
and describe radiation with negative energy density in terms of a suitable purely 
kinetic, non-canonical and ghost scalar field.
As we had mentioned in the introductory section, such ghost fields pose certain 
conceptual difficulties.
At this stage, we shall choose to overlook these difficulties and continue with 
our analysis.
We shall make a few remarks about the issue in the concluding section.

\par

Let the canonical field be $\phi$ and the non-canonical, ghost field be $\chi$.
We shall assume that the complete action describing these two fields is 
given by
\begin{equation}
S[\phi,\chi] 
= -\int \d^4 x\,\sqrt{-g}\,\l[- X^{^{\phi\phi}} + V(\phi)
+U_0\,\l({X^{^{\chi\chi}}}\r)^2\r],\label{eq:action}
\end{equation}
where $U_0$ is a positive constant, and the kinetic terms $X^{^{\phi\phi}}$ 
and $X^{^{\chi\chi}}$ are defined as
\begin{subequations}
\begin{eqnarray}
X^{^{\phi\phi}} &=& -\frac{1}{2}\,\partial_\mu\phi\,\partial^\mu\phi,\\
X^{^{\chi\chi}} &=& -\frac{1}{2}\,\partial_\mu\chi\,\partial^\mu\chi.
\end{eqnarray}
\end{subequations}
The stress-energy tensor associated with these fields can be obtained to be
\begin{subequations}
\begin{eqnarray}
T^{\mu}_{\nu(\phi)} 
&=& \partial^\mu\phi\,\partial_\nu\phi-
\delta^\mu_\nu\, \l[-X^{^{\phi\phi}}+V(\phi)\r],\\
T^{\mu}_{\nu(\chi)}
&=& -2\,U_0\,X^{^{\chi\chi}}\,\partial^\mu\chi\,\partial_\nu\chi
-\delta^\mu_\nu\,U_0\, \l(X^{^{\chi\chi}}\r)^2.
\end{eqnarray}
\end{subequations}

\par

Assuming the fields to be homogeneous, let us understand their behavior in a
bouncing universe.
Let us first consider the $\chi$ field. 
It is straightforward to obtain that 
\begin{subequations}
\begin{eqnarray}
T_{0(\chi)}^0 &=& -\rho_\chi \; = \frac{3\,U_0\,\dot{\chi}^4}{4}, \\
T_{j(\chi)}^i &=& p_\chi\,\delta_j^i = -\frac{U_0\,\dot{\chi}^4}{4}\,\delta_j^i.
\end{eqnarray}
\end{subequations}
Note that $\rho_{\chi}$ is negative and $p_{\chi}=\rho_{\chi}/3$, corresponding
to radiation.
In the absence of any potential, the equation of motion governing the field $\chi$
is extremely simple and is given by
\begin{equation}\label{eq:eom-chi}
\chi^{\prime\prime} = 0.
\end{equation}
This can be immediately integrated to obtain $\chi^\prime = C_2$, where $C_2$ 
is a constant of integration.
In other words, the field evolves monotonically towards either large or small
values as the universe evolves.
Such a behavior should not be surprising for a purely kinetic field that is
devoid of any potential to guide it.
The energy density $\rho_{\chi}$ can be written as
\begin{equation}
\rho_\chi = -\f{3\,U_0\,{\chi^\prime}^4}{4\,a^4}
= -\f{3\,U_0\, C_2^4}{4\,a^4},
\end{equation}
which is indeed the behavior of radiation, albeit with a negative energy density.

\par

Let us now turn to the behavior of the field $\phi$.
The components of the stress-energy tensor associated with the field are given by
\begin{subequations}\label{eq:rho-p-phi}
\begin{eqnarray}
T_{0(\phi)}^0 &=& -\rho_\phi = -\frac{\dot{\phi}^2}{2} - V\l(\phi\r),\\
T_{j(\phi)}^i &=& p_\phi\,\delta_j^i 
= \l[\frac{\dot{\phi}^2}{2} - V(\phi)\r]\,\delta_j^i.\label{eq:p-phi}
\end{eqnarray}
\end{subequations}
Recall that the field $\phi$ is expected to behave as ordinary matter.
The pressureless condition leads to [cf.~Eq.~(\ref{eq:p-phi})]
\begin{equation}
\f{\phi^{\prime\,2}}{2} - a^2\,V(\phi) = 0.\label{eq:v-p-phi}
\end{equation}
Further, being pressureless, the associated energy density is expected to
behave as, say, $\rho_\phi = C_1^2/a^3$, where $C_1$ is a constant.
This implies that we can can write [cf.~Eq.~(\ref{eq:rho-p-phi})]
\begin{equation}
\f{\phi^{\prime\,2}}{2} + a^2\,V(\phi) = \f{C_1^2}{a}.\label{eq:rho-phi} 
\end{equation}
Upon adding the above two equations, we obtain that
\begin{equation}
\phi^\prime = \f{C_1}{\sqrt{a}}.\label{eq:phi-p}
\end{equation}
Given the scale factor~(\ref{eq:sf}), this equation can be easily
integrated to arrive at
\begin{equation}
\phi = \phi_0\,\sinh^{-1}(k_0\,\eta),
\end{equation}
where $\phi_0 = C_1/(\sqrt{a_0}\, k_0)$ and we have set the constant
of integration to zero.
The above expression can be inverted to write
\begin{equation}
k_0\,\eta = \sinh\l(\frac{\phi}{\phi_0}\r).
\end{equation}
Since, according to Eqs.~(\ref{eq:v-p-phi}) and~(\ref{eq:phi-p}),
\begin{equation}
V(\phi)=\f{\phi'^2}{2\, a^2}=\f{C_1^2}{2\,a^3}, 
\end{equation}
on using the above solution for $\phi$, we can determine the potential to be
\begin{equation}
V(\phi) = \f{C_1^2}{2\,a_0^3\,\cosh^6\l(\phi/\phi_0\r)}.\label{eq:V}
\end{equation}
It is straightforward to check that the above expressions for the field and the 
potential indeed satisfy the following standard equation of motion governing 
the canonical scalar field:
\begin{equation}
\ddot{\phi} + 3\,H\,\dot{\phi} + V_{\phi} = 0,
\end{equation}
where $V_\phi=\d V/\d \phi$.
Note that the evolution of the field is symmetric about the bounce.
It starts with large negative values at early times during the contracting phase, 
rolls {\it up}\/ the potential~(\ref{eq:V}), reaching zero at the
bounce\footnote{The fact that the field rolls {\it up}\/ the potential during the 
contracting phase should not come as a surprise.
During an expanding phase such as inflation, $H$ is positive and, as is well
known, the $3\, H\, {\dot \phi}$ term leads to friction, slowing down the field that
is rolling down a potential.
In contrast, during a contracting phase, since $H$ is negative, the $3\, H\, 
{\dot \phi}$ term acts as `anti-friction', speeding up the field and thereby 
allowing it to climb the potential.}.
Thereafter, the field continues towards positive values, rolling down the potential
during the expanding phase.

\par

Now that we have arrived at the behavior of the fields, the remaining task is to
fix the constants $C_1$ and $C_2$.
They ought to be related to the parameters $a_0$ and $k_0$ in terms of which we
had expressed the scale factor and the constant $U_0$ that appears in the part of
the action describing the field $\chi$.
We find that the first Friedmann equation $3\,H^2\,\Mpl^2=\rho=\rho_\phi+\rho_\chi$
can be expressed as
\begin{equation}\label{eq:fix-coeff}
3\,\mathcal{H}^2\, \Mpl^2 
= \f{\phi^{\prime\,2}}{2} + a^2\,V(\phi) -\f{3\,U_0\,\chi'^4}{4\, a^2},
\end{equation}
where ${\cal H}=a'/a$ is the conformal Hubble parameter.
Upon using the various expressions we have obtained above and the scale 
factor~(\ref{eq:sf}), we can determine the constants $C_1$ and $C_2$ to be
\begin{subequations}
\begin{eqnarray}
C_1 &=& \sqrt{12\,a_0}\,\Mp\,k_0,\\
C_2 &=& \sqrt{\f{4\,\Mp\,a_0\,k_0}{U_0^{1/2}}},
\end{eqnarray}
\end{subequations}
so that the energy densities associated with the two fields reduce to
\begin{subequations}
\begin{eqnarray}
\rho_\phi &=& \f{12\,\Mp^2\,a_0\,k_0^2}{a^3},\\
\rho_\chi &=& -\f{12\,\Mp^2\,a_0^2\,k_0^2}{a^4}.
\end{eqnarray}
\end{subequations}
It is easy to see that $\rho_\phi + \rho_\chi = 0$ at the bounce, and such
a behavior would not have been possible without the ghost field $\chi$.

We should point out here that, if we make use the above expression for $C_1$,
the potential~(\ref{eq:V}) can be written as
\begin{equation}
V(\phi) = \f{6\,\Mpl^2\,(k_0/a_0)^2}{\cosh^6\l(\sqrt{12}\,\phi/\Mpl\r)}.
\end{equation}
In other words, the potential and, hence, the complete model, actually depends 
{\it only}\/ on the parameter $k_0/a_0$. 
Therefore, we can expect the power spectra to depend only on this combination.
This is already evident in the case of the tensors [cf. Eq.~(\ref{eq:ta})].
In due course, we shall see that similar conclusions apply to the scalars as 
well.
We shall comment further on this point in the concluding section.

\par

We should mention here that the matter bounce scenario driven by two scalar 
fields we are studying is somewhat similar to a system which had been  
investigated earlier~\cite{Allen:2004vz}.
In the earlier work, the purely kinetic, ghost field $\chi$ was described by a 
linear kinetic term, in contrast to the non-linear term that we are considering.
Also, the choice of the potential describing the canonical field $\phi$ was different. 
However, since both the models lead to a matter dominated phase at early times, 
we find that the two potentials behave in a similar manner at large negative 
values of the field.
The difference in the action governing the $\chi$ field and the choice of an even 
function for the potential describing the $\phi$ field lead to a difference in the
behavior of the background around the bounce between the two models.
Our choices not only permit us to solve for the background analytically, but, 
importantly, the symmetric matter bounce~(\ref{eq:sf}) of our interest leads 
to a tensor-to-scalar ratio that is consistent with the observations.


\section{Equations of motion governing the scalar 
perturbations}\label{sec:eom-sp}

In this section, we shall derive the equations governing the evolution of 
the scalar perturbations.
Since there are two fields involved, evidently, apart from the curvature 
perturbation, there will be an isocurvature perturbation present as well.
We shall derive the equations governing the perturbations $\delta\phi$ and 
$\delta\chi$ and their corresponding gauge invariant versions 
$\overline{\delta\phi}$ and $\overline{\delta\chi}$. 
Thereafter, we shall construct the curvature and isocurvature perturbations 
for our model and arrive at the corresponding equations governing them. 


\subsection{The first order Einstein's equations}

If we take into account the scalar perturbations to the background metric, 
then the FLRW line-element, in general, can be written as (see, for instance,
Refs.~\cite{Mukhanov:1990me,Martin:2004um,Bassett:2005xm,Sriramkumar:2009kg,
Sriramkumar:2012mik})
\begin{eqnarray}
{\rm d} s^2
= -\left(1+2\, A\right)\,{\rm d} t ^2 
+ 2\, a(t)\, (\partial_{i} B)\; {\rm d} t\; {\rm d} x^i
+a^{2}(t)\; \left[(1-2\, \psi)\; \delta _{ij}
+ 2\, \left(\partial_{i}\, \partial_{j}E \right)\right]\,
{\rm d} x^i\, {\rm d} x^j,\quad\;\label{eq:flrw-le-wsp}
\end{eqnarray}
where $A$, $B$, $\psi$ and $E$ are four scalar functions that describe 
the perturbations, which depend on time as well as space.
At the first order in the perturbations, the Einstein's equations are
given by~\cite{Mukhanov:1990me,Martin:2004um,Bassett:2005xm,Sriramkumar:2009kg,
Sriramkumar:2012mik}
\begin{subequations}\label{eq:fo-ee}
\begin{eqnarray}
3\,H\,\l(H\,A + \dot{\psi}\r) 
- \frac{1}{a^2}\,\nabla^2\l[\psi - a\,H\,\l(B - a\,\dot{E}\r)\r]
&=& -\frac{1}{2\,\Mp^2}\l(\delta\rho_\phi + \delta\rho_\chi\r),\\
\pa_i\l(H\,A + \dot{\psi}\r) 
&=& \frac{1}{2\,\Mp^2}\,\pa_i\l(\delta q_\phi + \delta q_\chi\r),\\
\ddot{\psi} + H\,\l(\dot{A} + 3\,\dot{\psi}\r) 
+ \l(2\,\dot{H} + 3\,H^2\r)\,A
&=& \frac{1}{2\,\Mp^2}\l(\delta p_\phi + \delta p_\chi\r),\\
A - \psi + \frac{1}{a}\l[a^2\,\l(B - a\,\dot{E}\r)\r]^{\cdot}&=&0
\end{eqnarray}
\end{subequations}
where $\delta\rho_I$ and $\delta p_I$, with $I=(\phi,\chi)$, are the perturbations 
in the energy densities and pressure associated with the two fields $\phi$ and $\chi$.
Further, the quantities $\delta q_I$ have been defined through the relation 
$\delta T^0_{i(I)}=-\pa_i(\delta q_I)$.
The last of the above equations follows from the fact that there are no anisotropic
stresses present.
The components of the perturbed stress-energy tensor of the two fields can be
evaluated to be
\begin{subequations}\label{eq:fo-set}
\begin{eqnarray}
\delta T_{0(\phi)}^0 
&=& -\delta\rho_\phi 
= -\dot{\phi}\,\dot{\delta\phi} + A\,\dot{\phi}^2 - V_{\phi}\,\delta\phi,\\
\delta T^0_{i(\phi)} 
&=& -\partial_i\,\delta q_\phi = -\partial_i\l(\dot{\phi}\,\delta\phi\r),\\
\delta T^i_{j(\phi)} 
&=& \delta p_\phi\,\delta^i_j 
= \l(\dot{\phi}\,\dot{\delta\phi} - A\,\dot{\phi}^2 
- V_{\phi}\,\delta\phi\right)\,\delta^i_j.
\end{eqnarray}
\end{subequations}
and
\begin{subequations}
\begin{eqnarray}
\delta T_{0(\chi)}^0 
&=& -\delta\rho_\chi 
= 3\,U_0\,\dot{\chi}^3\,\dot{\delta\chi} - 3\,A\,U_0\,\dot{\chi}^4,\\
\delta T^0_{i(\chi)} 
&=& -\partial_i\,\delta q_\chi 
= \partial_i\l(U_0\,\dot{\chi}^3\,\delta\chi\r),\\
\delta T^i_{j(\chi)} 
&=& \delta p_\chi\,\delta^i_j = \l(U_0\,A\,\dot{\chi}^4 
- U_0\,\dot{\chi}^3\,\dot{\delta\chi}\r)\,\delta^i_j.
\end{eqnarray}
\end{subequations}


\subsection{Equations governing the perturbations in the scalar fields}

The equations of motion describing the perturbations in the fields can 
be arrived at from the following conservation equation governing the 
perturbation in the energy density of a specific component (see, 
for instance, Refs.~\cite{Malik:2004tf,Malik:2008im}):
\begin{equation}
\dot{\delta \rho}_I
+3\, H\, \l(\delta \rho_I+\delta p_I\r)
-3\,(\rho_I+p_I)\,\dot{\psi}
-\nabla^2\l[\l(\f{\rho_I+p_I}{a}\r)\,B +\f{\delta q_I}{a^2}
-(\rho_I+p_I)\,\dot{E}\r]=0.
\end{equation}
Upon making use of this equation and the above expressions for the components
of the perturbed stress-energy tensor, we can obtain the equations of motion 
governing the Fourier modes, say, $\delta\phi_k$ and $\delta\chi_k$, 
associated with the perturbations in the two scalar fields to be
\begin{subequations}\label{eq:eom-delta-phi-chi}
\begin{eqnarray}
\ddot{\delta\phi}_k + 3\,H\,\dot{\delta\phi}_k + V_{\phi\phi}\,\delta\phi_k
+ 2\,V_{\phi}\,A_k\!\! 
& &-\, \dot{\phi}\,\l(\dot{A}_k + 3\,\dot{\psi}_k\r)\nn\\
& &+\,\,\,\,\frac{k^2}{a^2}\,\l[\delta\phi_k +a\,\dot{\phi}\,
\l(B_k-a\,\dot{E}_k\r)\r]= 0,\label{eq:eom-delta-phi}\qquad\;\;\\
\ddot{\delta\chi}_k + H\, \dot{\delta\chi}_k 
- \dot{\chi}\,\!\l(\dot{A}_k + \dot{\psi}_k\r)\!\!
& &+\,\frac{k^2}{3\,a^2}\,\!\l[\delta\chi_k+a\,\dot{\chi}\,
\l(B_k-a\,\dot{E}_k\r)\r] 
= 0,\label{eq:eom-delta-chi}\qquad
\end{eqnarray}
\end{subequations}
where, evidently, $A_k$, $B_k$, $\psi_k$ and $E_k$ denote the Fourier modes 
that describe the corresponding metric perturbations.
The gauge invariant perturbations associated with the two scalar fields can
be constructed to be
\begin{subequations}
\begin{eqnarray}
\overline{\delta\phi}_k &=& \delta\phi_k 
+ \frac{\dot{\phi}}{H}\,\psi_k,\\
\overline{\delta\chi}_k 
&=& \delta\chi_k + \frac{\dot{\chi}}{H}\,\psi_k.
\end{eqnarray}
\end{subequations}
Upon using the equations of motion~(\ref{eq:eom-delta-phi-chi}) and the 
first order Einstein equations~(\ref{eq:fo-ee}), we find that these 
gauge invariant perturbations of the two scalar fields obey the following 
equations:
\begin{subequations}\label{eq:eom-gi-delta-phi-chi}
\begin{eqnarray}
& &\overline{\delta\phi}_k'' + 2\,\cH\,\overline{\delta\phi}_k' 
+ \l(k^2 + a^2\,V_{\phi\phi} + \frac{2\,a^2\,\phi'\,V_\phi}{\cH\,\Mp^2} 
+ \f{3\,\phi'^2}{\Mp^2} - \frac{\phi'^4}{2\,\cH^2\,\Mp^4} 
+ \f{U_0\,\phi'^2\,\chi'^4}{a^2\,\cH^2\,\Mp^4}\r)\,\overline{\delta\phi}_k\nn\\
& &\qquad
=\,\f{U_0\,\phi'\,{\chi'}^3}{a^2\,\cH\,\Mp^2}\,\overline{\delta\chi}_k' 
+ \l(\f{U_0\,V_\phi\,\chi'^3}{\cH\,\Mp^2} 
+ \frac{3\,U_0\,\phi'\,\chi'^3}{a^2\,\Mp^2} 
- \frac{U_0\,{\phi'}^3\,\chi'^3}{2\,a^2\,\cH^2\,\Mp^4} 
+ \frac{U_0^2\,\phi'\,\chi'^7}{a^4\,\cH^2\,\Mp^4}\right)\overline{\delta\chi}_k,\qquad\\
& &\overline{\delta\chi}_k'' 
+ \l(\f{k^2}{3} - \f{2\,U_0\,\chi'^4}{a^2\,\Mp^2}
+ \f{U_0\,{\phi'}^2\,\chi'^4}{3\,a^2\,\cH^2\,\Mp^4} 
- \f{U_0^2\,\chi'^8}{2\,a^4\,\cH^2\,\Mp^4}\right)\,\overline{\delta\chi}_k\nn\\ 
& &\qquad 
= \f{\phi'\,\chi'}{3\,\cH\,\Mp^2}\,\overline{\delta\phi}_k'
- \l(\f{2\,\chi'\,a^2\,V_\phi}{3\,\cH\,\Mp^2} 
+ \f{2\,\phi'\,\chi'}{\Mp^2} 
- \f{{\phi'}^3\,\chi'}{3\,\cH^2\,\Mp^4} 
+ \f{U_0\,\phi'\,\chi'^5}{2\,a^2\,\cH^2\,\Mp^4}\r)\,\overline{\delta\phi}_k.
\end{eqnarray}
\end{subequations}
Let us now turn to the construction of the gauge invariant curvature and
isocurvature perturbations associated with the two fields.
In due course, we shall make use of the above equations to obtain the equations
governing the curvature and isocurvature perturbations.


\subsection{Constructing the curvature and isocurvature 
perturbations}\label{subsec:curv-isocurv}

As is well known, in the presence of more than one field or fluid, apart from 
the curvature perturbation, isocurvature perturbations are also generated. 
The isocurvature perturbations source the curvature perturbations.
It is the structure of the complete action describing the matter fields that 
determines the relation between the perturbations in the fields and the 
curvature and isocurvature perturbations. 
While the fluctuations along the direction of the background trajectory in the 
field space are referred to as the adiabatic or the curvature perturbation, the 
perturbations along a direction perpendicular to the background trajectory are 
called the non-adiabatic, entropic or isocurvature 
perturbations~\cite{Gordon:2000hv,Malik:2004tf,Malik:2008im}.

\par

The Lagrangian density associated with the action~(\ref{eq:action}) is evidently
given by
\begin{equation}
\cL = X^{^{\phi\phi}} - V(\phi)-U_0\left(X^{^{\chi\chi}}\right)^2.
\end{equation}
Let us now define a set of basis vectors along the direction of background evolution, 
\viz the adiabatic basis, and another set of basis vectors along the direction 
perpendicular to the background evolution, which is referred to as the entropic basis. 
These two sets of basis vectors obey the following orthonormality condition (see, for 
instance, Refs.~\cite{Langlois:2008mn,Langlois:2008qf}):
\begin{equation}
\cL_{_{X^{^{IJ}}}}\,e_n^I\,e_m^J = \delta_{nm},\label{eq:on}
\end{equation}
where $(I,J)=(\phi,\chi)$, $(n,m)=(1,2)$ and
\begin{equation}
\cL_{_{X^{^{IJ}}}}=\f{\pa \cL}{\pa X^{^{IJ}}}.
\end{equation}
The adiabatic basis vectors can be defined as \cite{Langlois:2008mn,Langlois:2008qf}
\begin{equation}
e_1^I 
= \f{\dot{\varphi}^I}{\sqrt{\cL_{_{X^{^{JK}}}}\,\dot{\varphi}^J\,\dot{\varphi}^K}},
\end{equation}
where $(\varphi^1,\varphi^2)=(\phi,\chi)$\footnote{Actually, since $I$ already 
represents $\phi$ and $\chi$, the introduction of $\varphi^I$ implying 
$(\varphi^1,\varphi^2)=(\phi,\chi)$ may be considered as redundant notation.
However, representing the perturbations in the scalar fields as $\delta\varphi^I
=(\delta\phi,\delta \chi)$ seems to be a better choice than denoting them as
$\delta I$!}.
Since 
\begin{equation}
\cL_{_{X^{^{\phi\phi}}}} = 1
\end{equation}
and
\begin{equation}
\cL_{_{X^{^{\chi\chi}}}} 
= -2\,U_0\,X^{^{\chi\chi}} = -U_0\,\dot{\chi}^2,
\end{equation}
we can define the two adiabatic basis vectors to be
\begin{subequations}\label{eq:basis-R}
\begin{eqnarray}
e_1^\phi &=& \f{\dot{\phi}}{\sqrt{\dot{\phi}^2 - U_0\,\dot{\chi}^4}},\\
e_1^\chi &=& \f{\dot{\chi}}{\sqrt{\dot{\phi}^2 - U_0\,\dot{\chi}^4}}.
\end{eqnarray}
\end{subequations}
The curvature perturbation can be defined in terms of these basis vectors as
\begin{eqnarray}
\cR = \f{H}{\cL_{_{X^{^{IJ}}}}\dot{\varphi}^I\,\dot{\varphi}^J}\;
\cL_{_{X^{^{KL}}}}\,\dot{\varphi}^K\,\overline{\delta\varphi}^L 
= \f{H}{\sqrt{\cL_{_{X^{^{IJ}}}}\dot{\varphi}^I\,\dot{\varphi}^J}}\;
\cL_{_{X^{^{KL}}}}\,e_1^K\,\overline{\delta\varphi}^L,
\end{eqnarray}
where $\overline{\delta\varphi}^L$ is the gauge invariant perturbation associated
with the field $\varphi^L$.
For our model, the curvature perturbation can be constructed to be
\begin{equation}
\cR 
= \f{H}{\dot{\phi}^2 - U_0\,\dot{\chi}^4}\,\l(\dot{\phi}\,\overline{\delta\phi}
- U_0\,\dot{\chi}^3\,\overline{\delta\chi}\right).\label{eq:cR}
\end{equation}
It is well known that, when multiple components (fluids and/or fields) are present, 
the total curvature perturbation is defined as (see, for instance, 
Refs.~\cite{Malik:2004tf,Malik:2008im})
\begin{equation}
\cR 
= \sum_{I}\f{\rho_I+p_I}{\rho+p}\, \cR_I,\label{eq:cR-gd}
\end{equation}
where $\cR_I$ is the curvature perturbation associated with an individual component
and is given by
\begin{equation}
\cR_I=\psi+\f{H}{\rho_I+p_I}\, \delta q_I.\label{eq:cRI}
\end{equation}
For our system, it is easy to show that, if we make use of the expressions 
for the various quantities we have obtained earlier, the definition~(\ref{eq:cR-gd}) 
for the total curvature perturbation indeed matches the explicitly gauge invariant
expression~(\ref{eq:cR}) we have arrived at.
Note that the expression~(\ref{eq:cR}) for $\cR$ suggests that it may diverge when 
$\dot{\phi}^2 - U_0\,\dot{\chi}^4 = 0$, which corresponds to the condition $\dot{H} 
= 0$. 
Recall that, $\dot{H}=0$ at $\mp\eta_\ast = \mp \eta_0/\sqrt{3}$.
As we shall see, the curvature perturbation indeed diverges at these times.
The expression~(\ref{eq:cR}) also suggests that the curvature perturbation may turn
out to be zero at the bounce, wherein $H=0$.
However, we find that this actually does not occur at the bounce, but the curvature 
perturbation vanishes for an instant between the bounce and $\eta_\ast$.

\par 

Let us now construct the corresponding basis vectors for the entropic perturbations, 
\viz $e_2^\phi$ and $e_2^\chi$.
Using Eqs.~(\ref{eq:basis-R}) and the orthonormality condition~(\ref{eq:on}), we obtain
that
\begin{subequations}\label{eq:basis-S}
\begin{eqnarray}
e_2^\phi 
&=& \f{\dot{\chi}\,\sqrt{-U_0\,\dot{\chi}^2}}{\sqrt{\dot{\phi}^2 
- U_0\,\dot{\chi}^4}},\\
e_2^\chi 
&=& \f{\dot{\phi}}{\sqrt{-U_0\,\dot{\chi}^2}\sqrt{\dot{\phi}^2
- U_0\,\dot{\chi}^4}}.
\end{eqnarray}
\end{subequations}
It is straightforward to check that these two basis vectors are indeed 
orthogonal to the direction of the background evolution. 
The isocurvature perturbation can therefore be defined in terms of the basis 
vectors~(\ref{eq:basis-S}) as
\begin{equation}
\cS = \f{H}{\sqrt{\cL_{_{X^{^{IJ}}}}\dot{\varphi}^I\,\dot{\varphi}^J}}\;
\cL_{_{X^{^{KL}}}}\,e_2^K\,\overline{\delta\varphi}^L.
\end{equation}
This can be expressed as
\begin{equation}
\cS = \f{H\,\sqrt{U_0\,\dot{\chi}^2}}{\dot{\phi}^2 - U_0\,\dot{\chi}^4}\,
\l(\dot{\chi}\, \overline{\delta\phi} - \dot{\phi}\,\overline{\delta\chi}\r),
\end{equation}
where, in order for $\cS$ to remain a real quantity, we have dropped the
minus sign under the square root that appears as an overall coefficient.
It is easy to check that, apart from an overall background factor, the 
isocurvature perturbation we have defined above can be expressed as the
difference of the curvature perturbation $\cR_I$ [cf.~Eq.~(\ref{eq:cRI})] 
associated with the two individual fields~\cite{Malik:2004tf}.
Note that, as in the case of the curvature perturbation, the isocurvature 
perturbation can be expected to diverge at $\mp\eta_\ast$ and vanish at 
the bounce.
We shall see later that these expectations indeed prove to be true.


\subsection{Equations governing the curvature and the 
isocurvature perturbations}\label{subsec:curv-isocurv-eqns}

Let $\cR_k$ and $\cS_k$ denote the Fourier modes associated with the curvature 
and the isocurvature perturbations.
The expressions for the curvature and the isocurvature perturbations we have
obtained above can be inverted to arrive at the following relations:
\begin{subequations}
\begin{eqnarray}
\overline{\delta\phi}_k 
&=& \f{1}{\cH}\,\l(\phi'\,\cR_k - \f{1}{a}\,\sqrt{U_0\,\chi'^4}\,\cS_k\r),\\
\overline{\delta\chi}_k 
&=& \f{1}{\cH}\,\l(\chi'\,\cR_k 
- \frac{a\,\phi'}{\sqrt{U_0\,\chi'^2}}\,\cS_k\r).
\end{eqnarray}
\end{subequations}
Using the equations of motion for the gauge invariant field 
perturbations~(\ref{eq:eom-gi-delta-phi-chi}), we obtain the 
equations governing $\cR_k$ and $\cS_k$ to be
\begin{subequations}
\begin{eqnarray}
\cR_k''
&+& \biggl\{\f{2\,\l(\cH'-\cH^2\r)}{\cH}-2\,\cH 
+ \f{a}{\Mp^2}\,\l[\frac{2\,\phi'^2}{3\,a\,\cH} 
- \f{a\,V_\phi\,\phi'}{\cH^\prime-\cH^2} 
- \f{\cH\,\phi'^2}{a\, \l(\cH'-\cH^2\r)}\r]\nn\\ 
& &-\, \f{\phi'^4}{3\,\Mp^4\,\cH\,\l(\cH^\prime-\cH^2\r)}\biggr\}\,\cR_k'
+ \f{k^2}{3}\,\l[1 + \f{\phi'^2}{\Mp^2\,\l(\cH^\prime-\cH^2\r)}\r]\,\cR_k\nn\\
&=& \f{\sqrt{U_0\,\chi'^4}\,\phi'}{\Mp^2\,\l(\cH'-\cH^2\r)}\,
\l[\f{a\,V_\phi}{\phi'} + \f{\cH}{a} 
- \f{1}{2\,\Mp^2\,\cH}\,
\l(\f{\phi'^2}{a}- \f{U_0\,\chi'^4}{3\,a^3}\r)\r]\,\cS_k'\nn\\
& &+\, \f{\sqrt{U_0\,\chi'^4}\,\phi'}{a\,\Mpl^2}\,
\biggl[\f{k^2}{3\,\l(\cH^\prime-\cH^2\r)}
+ \f{5}{3} + \f{5\,a^2\,V_\phi}{\cH\,\phi'} 
- \f{2\,\cH^2}{\cH'-\cH^2}
+ \f{V_{\phi\phi}\,a^2}{\cH^\prime-\cH^2} 
- \f{\cH^\prime-\cH^2}{3\,\cH^2}\nn\\
& &+\, \f{1}{\Mp^2}\,\l(\frac{2\,V_\phi\,\phi'\,a^2}{3\,\cH\,\l(\cH^\prime-\cH^2\r)} 
+ \f{\phi'^2}{\cH^\prime-\cH^2} 
- \f{\phi'^2}{3\,\cH^2}\r)\biggr]\,\cS_k,\\
\cS_k'' 
&+&\!\! \l[\f{2\,\l(\cH'-\cH^2\r)}{\cH}
-2\,\cH - \f{1}{\Mp^2}\,\!\!\l(\f{2\,\phi'^2}{3\,\cH} 
+ \f{V_\phi\,\phi'\,a^2}{\cH'-\cH^2} 
+ \f{\cH\,\phi'^2}{\cH'-\cH^2}\r) 
+ \frac{\phi'^4}{3\,\Mp^4\,\cH\,\l(\cH^\prime-\cH^2\r)}\r]\,\!\!\cS_k'\nn\\
&+& \biggl\{k^2\,\l[1 - \f{\phi'^2}{3\,\Mp^2\,\l(\cH'-\cH^2\r)}\r] 
- 2\,\cH^2 + a^2\,V_{\phi\phi} + \f{2\,\l(\cH'-\cH^2\r)^2}{\cH^2} 
- 3\,\l(\cH'-\cH^2\r) \nn\\
& &+\, \f{1}{\Mp^2}\,
\l[\f{2\,\cH^2\,\phi'^2}{\cH'-\cH^2} -\f{2\,V_\phi\,\phi'\,a^2}{3\,\cH} 
- \f{2\,\phi'^2}{3}- \f{V_{\phi\phi}\,\phi'^2\,a^2}{\cH'-\cH^2}
- \f{2\,\phi'^2\,\l(\cH'-\cH^2\r)}{3\,\cH^2}\r]\nn\\ 
& &+\, \f{1}{\Mp^4}\,\l[\f{\phi'^4}{3\,\cH^2}
- \f{2\,V_\phi\,\phi'^3\,a^2}{3\,\cH\,\l(\cH'-\cH^2\r)}
-\f{2\,\phi'^4}{3\,\l(\cH'-\cH^2\r)}\r]\biggr\}\,\cS_k\nn\\
&=& \f{\sqrt{U_0\,\chi'^4}\,\phi'}{\Mp^2\,\l(\cH'-\cH^2\r)}\,\!\!
\l(\f{\cH'-\cH^2}{a\,\cH} - \f{\cH}{a} - \f{a\,V_\phi}{\phi'} 
- \f{\phi'^2}{3\,\Mp^2\,a\,\cH}\r)\,\cR_k'
+ \f{\sqrt{U_0\,\chi'^4}\,\phi'}{3\,a\,\Mp^2\,\l(\cH'-\cH^2\r)}\,k^2\,\cR_k.\nn\\
\end{eqnarray}
\end{subequations}
We should stress here that these equations apply to the two field model 
described by the action~(\ref{eq:action}). 
For the specific bouncing scenario of our interest characterized by the 
scale factor~(\ref{eq:sf}), these equations simplify to be
\begin{subequations}\label{eq:eom-cRk-cSk-ss}
\begin{eqnarray}
& &\!\!\!\!
\cR_k'' + \f{2\,\l(7 + 9\,k_0^2\,\eta^2 - 6\,k_0^4\,\eta^4\r)}{\eta\, 
\l(1 - 3\,k_0^2\,\eta^2\r)\,\l(1 +k_0^2\,\eta^2\r)}\,\cR_k'
- \f{k^2\,\l(5 + 9\,k_0^2\,\eta^2\r)}{3\, \l(1- 3\,k_0^2\,\eta^2\r)}\,\cR_k  \nn\\
& &\;\;\;
= \f{4\l(5 + 12\,k_0^2\,\eta^2\r)}{\sqrt{3}\,\eta\,\l(1 - 3\,k_0^2\,\eta^2\r)
\sqrt{1+ k_0^2\,\eta^2}}\,\cS_k' 
- \f{4\,\l[5 - 22\,k_0^2\,\eta^2 - 24\,k_0^4\,\eta^4 + k^2\,\eta^2
\l(1 + k_0^2\,\eta^2\r)^2\r]}{\sqrt{3}\,\eta^2\,\l(1 + k_0^2\,\eta^2\r)^{3/2}
\l(1- 3\,k_0^2\,\eta^2\r)}\,\cS_k,\nn\\
\label{eq:eom-cRk-ss}\\
& &\!\!\!\!
\cS_k''- \f{2\,\l(9 + 7\,k_0^2\,\eta^2 
+ 6\,k_0^4\,\eta^4\r)}{\eta\,\l(1 - 3\,k_0^2\,\eta^2\r)\,
\l(1 +\,k_0^2\,\eta^2\r)}\,\cS_k^\prime\nn\\
& &\quad\!\!
+\, \frac{18 - 85\,k_0^2\,\eta^2 - 25\,k_0^4\,\eta^4 - 6\,k_0^6\,\eta^6 
+ k^2\,\eta^2\,\l(3 - k_0^2\,\eta^2\r)\,
\l(1+ k_0^2\,\eta^2\r)^2}{\eta^2\,\l(1 - 3\,k_0^2\,\eta^2\r)\,
\l(1 + k_0^2\,\eta^2\r)^2}\,\cS_k \nn\\
& &\quad\!\!
=\, -\f{4\,\sqrt{3}\,\l(3 - 2\,k_0^2\,\eta^2\r)}{\eta\,\sqrt{1 + k_0^2\,\eta^2}
\l(1 - 3\,k_0^2\,\eta^2\r)}\,\cR_k'
+ \f{4\,k^2\,\sqrt{1 + k_0^2\,\eta^2}}{\sqrt{3}\,\l(1 - 3\,k_0^2\,\eta^2\r)}\,\cR_k.
\label{eq:eom-cSk-ss}
\end{eqnarray}
\end{subequations}
Note that the denominators of some of the coefficients in these equations 
contain either a factor of $\eta$ or $(1-3\, k_0^2\, \eta^2)$. 
Therefore, as one approaches the bounce during the contracting phase, the 
coefficients will first diverge at $-\eta_\ast$ and then at the bounce. 
Later, after the bounce, they will also diverge at $\eta_\ast$, before we 
get to evaluate the power spectra.
Due to this reason, the above equations do not permit us to evolve the
quantities $\cR_k$ and $\cS_k$ across the bounce.
This issue can be circumvented by working in a specific gauge and considering
two other suitable quantities to characterize the perturbations whose governing 
equations remain well behaved around the bounce (see Ref.~\cite{Allen:2004vz},
in this context, also see Ref.~\cite{Battefeld:2004mn}).

Another related point needs to be emphasized at this stage of our discussion.
As we shall describe in some detail in the next section, the initial conditions 
on the perturbations need to be imposed at sufficiently early times when the
modes are well inside the Hubble radius during the contracting phase.
Moreover, in order to impose the standard initial conditions on the curvature and
isocurvature perturbations, the modes need to be decoupled during these early
times.
It has been pointed out that a strong coupling between the two modes would 
not permit the imposition of standard, independent initial conditions on the 
modes (for a detailed discussion on this issue, see Ref.~\cite{Peter:2015zaa}).
In due course, we shall discuss the specific initial conditions that we shall 
impose on the perturbations (see Sub-sec.~\ref{subsec:ic}).
We ought to stress here that the Eqs.~(\ref{eq:eom-cRk-cSk-ss})
governing $\cR_k$ and $\cS_k$ indeed decouple at very early times, \ie~as 
$\eta \to -\infty$ [cf. Eqs.~(\ref{eq:eom-cRk-d1}) and (\ref{eq:eom-cSk-d1})].
We should highlight the fact that, in the next two sections, apart from the 
numerical solutions, we shall also construct analytical solutions, which we 
shall show match the numerical results very well.


\subsection{Perturbations in a specific gauge}\label{subsec:uniform-chi}

We now need to identify a suitable gauge wherein the perturbations can be 
evolved across the bounce without facing the difficulties mentioned above.
We find that these difficulties can be avoided if we choose to work in the 
uniform-$\chi$ gauge~\cite{Allen:2004vz}. 
In this gauge, the two independent scalar perturbations turn out to be the 
metric potentials $A$ and $\psi$ and, as we shall soon illustrate, these 
quantities can be smoothly evolved across the bounce.
The curvature and the isocurvature perturbations can then be suitably 
constructed from these two scalar perturbations.

\par

The uniform $\chi$-gauge corresponds to the situation wherein $\delta\chi_k = 0$. 
In such a case, Eq.~(\ref{eq:eom-delta-chi}) reduces to
\begin{equation}\label{eq:eom-delta-chi-0}
\frac{k^2}{3\,a}\,\l(B_k-a\,\dot{E}_k\r) 
= \l(\dot{A}_k + \dot{\psi}_k\r).
\end{equation}
Upon using this relation, the first order Einstein equations~(\ref{eq:fo-ee}) and
the background equations, we obtain the following equations governing $A_k$ and 
$\psi_k$:
\begin{subequations}\label{eq:eom-Ak-psik-eta}
\begin{eqnarray}
A_k'' + 4\,\mathcal{H}\,A_k'
+\biggl[\f{k^2}{3} - \biggl(6\,\cH^2 &-& \f{\phi'^2}{\Mp^2}
+\f{2\,a^2\,\cH\, V_\phi}{\phi'}
+\f{2\,U_0\,\chi'^4}{a^2\,\Mp^2}\biggr)\biggr]\,A_k \\ \nn
&=& \f{2\,a^2\,V_\phi}{\phi'}\,\psi_k' 
+ \frac{4\,k^2}{3}\,\psi_k,\label{eq:eom-Ak-eta}\\
\psi_k'' 
+\l(2\,\cH +\f{2\,a^2\, V_\phi}{\phi'}\r)\,\psi_k' + k^2\, \psi_k 
&=& 2\,\mathcal{H}\,A_k'-\biggl(6\,\cH^2
 - \f{\phi'^2}{\Mp^2}
+\f{2\,a^2\,\cH\, V_\phi}{\phi'}
+\f{2\,U_0\,\chi'^4}{a^2\,\Mp^2}\biggr)\,A_k.
\label{eq:eom-psik-eta}\nn\\
\end{eqnarray}
\end{subequations}
We should again mention that these equations correspond to the system described
by the action~(\ref{eq:action}).
For the specific bouncing scenario that we are considering here, the above equations
simplify to
\begin{subequations}\label{eq:eom-Ak-psik-eta-ss}
\begin{eqnarray}
A_k'' + 4\,\mathcal{H}\,A_k'
+\l(\frac{k^2}{3} - \f{20\,a_0^2\, k_0^2}{a^2}\r)\,A_k 
&=&- 3\,\mathcal{H}\,\psi_k' + \frac{4\,k^2}{3}\,\psi_k,\label{eq:eom-Ak-eta-ss}\\
\psi_k'' - \mathcal{H}\,\psi_k' + k^2\, \psi_k 
&=& 2\,\mathcal{H}\,A_k'-\frac{20\,a_0^2\, k_0^2}{a^2}\,A_k.
\label{eq:eom-psik-eta-ss}
\end{eqnarray}
\end{subequations}
In arriving at these two equations, we have made use of the 
relation:~$\dot{\phi}^2/2 = V\l(\phi\r)$, which arises due to the fact that 
the field $\phi$ is pressureless.
Note that, in the uniform $\chi$-gauge, the curvature and isocurvature perturbations 
are given by
\begin{subequations}\label{eq:RS-Apsi}
\begin{eqnarray}
\cR_k &=& \psi_k 
+ \frac{2\,H\,\Mpl^2}{\dot{\phi}^2 - U_0\,\dot{\chi}^4}\,
\l(\dot{\psi}_k+H\, A_k\r),\\
\cS_k &=& \frac{2\,H\,\Mpl^2\,\sqrt{U_0\,\dot{\chi}^4}}{\l(\dot{\phi}^2 
- U_0\,\dot{\chi}^4\r)\,\dot{\phi}}\, \l(\dot{\psi}_k+H\, A_k\r).
\end{eqnarray}
\end{subequations}
Later, we shall make use of these relations to construct $\cR_k$ and $\cS_k$
from $A_k$ and $\psi_k$ around the bounce.


\section{Evolution of the scalar perturbations}\label{sec:n}

Let us now turn to solving the equations governing the scalar perturbations
numerically.
Since we have analytical solutions to describe the behavior of the background 
quantities, we need to develop the numerical procedure only for the evolution 
of the perturbations. 
Our main aim is to evaluate the scalar power spectra after the bounce, which,
obviously, requires us to evolve the perturbations across the bounce.
In the case of tensors, we could evolve the perturbations smoothly across the 
bounce and evaluate the corresponding power spectrum at a suitable time after 
the bounce.
However, in the case of scalars, as we have described above, it does not seem
possible to integrate the equations describing the curvature and the isocurvature 
perturbations across the bounce due to the presence of diverging coefficients.
We shall hence choose to evolve the metric perturbations $A_k$ and $\psi_k$ 
across the bounce, since the equations governing them are devoid of such divergent 
terms.
Once we have evolved $A_k$ and $\psi_k$ across the bounce, we shall reconstruct 
the curvature and the isocurvature perturbations $\cR_k$ and $\cS_k$ from these
quantities to arrive at the power spectra.

\par

Recall that, in the case of tensors, when evaluating the perturbations 
{\it analytically},\/ we had divided the period of our interest -- \ie 
from very early times during the contracting phase to a suitable time 
immediately after the bounce -- into two domains, \viz $-\infty < \eta 
\le -\alpha\, \eta_0$ and $-\alpha\, \eta_0 \le \eta \le \beta\, \eta_0$, 
where we had set $\alpha = 10^5$ and $\beta = 10^2$.
In the case of scalars, we shall work over these two domains to evolve the 
perturbations {\it analytically as well as numerically}.\/
In the first domain, we shall identify the Mukhanov-Sasaki variables associated
with the perturbations $\cR_k$ and $\cS_k$ and impose the corresponding Bunch-Davies 
initial conditions on these variables at suitably early times.
We shall evolve the  perturbations $\cR_k$ and $\cS_k$ using the governing
equations~(\ref{eq:eom-cRk-cSk-ss}) until $\eta=-\alpha\, \eta_0$.
At $\eta=-\alpha\, \eta_0$, we shall match the quantities $\cR_k$ and $\cS_k$
(and their time derivatives) to the metric perturbations $A_k$ and $\psi_k$ 
(and their time derivatives) using the relations~(\ref{eq:RS-Apsi}).
Thereafter, we shall evolve the perturbations $A_k$ and $\psi_k$
[using Eqs.~(\ref{eq:eom-Ak-psik-eta-ss})] until $\eta=\beta\, \eta_0$ after 
the bounce.
Once we have evolved $A_k$ and $\psi_k$ across the bounce, we can reconstruct 
the quantities $\cR_k$ and $\cS_k$ [using Eqs.~(\ref{eq:RS-Apsi})] and also,
eventually, evaluate their power spectra.


\subsection{Equations in terms of e-$\cN$-folds}

As in the case of tensors, we shall numerically integrate the equations with 
e-$\cN$-folds as the independent variable.
We need to numerically integrate the equations governing the evolution of the 
quantities $\cR_k$, $\cS_k$, $\psi_k$ and $A_k$.
In terms of the variable e-$\cN$-folds, Eqs.~(\ref{eq:eom-cRk-cSk-ss}) can be 
written as
\begin{subequations}\label{eq:eom-cRk-cSk-ss-cN}
\begin{eqnarray}
\f{\d^2 \cR_k}{\d\cN^2} 
&+& \l[\cN+\f{1}{H}\,\f{\d H}{\d\cN} - \frac{1}{\cN} 
+ \f{2\,a_0\,\cN}{a^2\,H\, \eta}
\l(\frac{7 + 9\,k_0^2\,\eta^2 - 6\,k_0^4\,\eta^4}{1 - 3\,k_0^2\,\eta^2}\r)\r]\,
\f{\d \cR_k}{\d\cN}\nn\\
&-& \f{k^2\,\cN^2}{3\,a^2\,H^2}\,
\l(\f{5 + 9\,k_0^2\,\eta^2}{1 - 3\,k_0^2\,\eta^2}\r)\,\cR_k\nn\\
&=& \f{4\,a_0^{1/2}\,\cN}{\sqrt{3}\,a^{3/2}\,H\,\eta}\,
\l(\f{5 + 12\,k_0^2\,\eta^2}{1 - 3\,k_0^2\,\eta^2}\r)\,\f{\d \cS_k}{\d\cN}\nn\\
& &-\,\f{4\,\cN^2\,a_0^{3/2}}{\sqrt{3}\,a^{7/2}\,H^2\,\eta^2}\,
\l[\f{5 - 22\,k_0^2\,\eta^2 - 24\,k_0^4\,\eta^4 
+ k^2\,\eta^2\,\l(a/a_0\r)^2}{1- 3\,k_0^2\,\eta^2}\r]\,\cS_k,\label{eq:eom-cR-N}\\ 
\f{\d^2 \cS_k}{\d\cN^2}  
&+& \l[\cN+\f{1}{H}\,\f{\d H}{\d\cN}- \frac{1}{\cN} 
- \f{2\,a_0\,\cN}{a^2\,H\, \eta}\,
\l(\f{9 + 7\,k_0^2\,\eta^2 + 6\,k_0^4\,\eta^4}{1 - 3\,k_0^2\,\eta^2}\r)\r]\,
\f{\d \cS_k}{\d\cN}\nn\\
&+& \f{a_0^2\,\cN^2}{a^4\,H^2\,\eta^2}
\l(\f{18 - 85\,k_0^2\,\eta^2 - 25\,k_0^4\,\eta^4 
- 6\,k_0^6\,\eta^6 + k^2\,\eta^2\,\l(3 - k_0^2\,\eta^2\r)\,
\l(a/a_0\r)^2}{1 - 3\,k_0^2\,\eta^2}\r)\,\cS_k\nn\\
&=& -\f{4\,\sqrt{3 }\,a_0^{1/2}\,\cN}{a^{3/2}\, H\, \eta}\,
\l(\f{3 - 2\,k_0^2\,\eta^2}{1 - 3\,k_0^2\,\eta^2}\r)\,\f{\d \cR_k}{\d\cN}
+ \f{4\,k^2\,\cN^2}{\sqrt{3}\, a_0^{1/2}\,a^{3/2}\,H^2}\,
\l(\f{1}{1 - 3\,k_0^2\,\eta^2}\r)\,\cR_k.\label{eq:eom-cS-N}
\end{eqnarray}
\end{subequations}
Similarly, we find that Eqs.~(\ref{eq:eom-Ak-psik-eta-ss}) can be expressed as
\begin{subequations}\label{eq:eom-Ak-psik-cN}
\begin{eqnarray}
\f{\d^2 A_k}{\d\cN^2}  
+ \l(5\,\cN+\f{1}{H}\,\f{\d H}{\d\cN}- \f{1}{\cN} \r)\,\f{\d A_k}{\d\cN}
&+& \l(\f{k^2\,\cN^2}{3\,a^2\,H^2}
- \f{20\,a_0^2\,\cN^2\,k_0^2}{a^4\,H^2}\r)\,A_k\nn\\ 
&=&-\,3\,\cN\,\f{\d \psi_k}{\d\cN} 
+ \f{4\,k^2\,\cN^2}{3\,a^2\,H^2}\,\psi_k,\label{eq:eom-Ak-cN}\\
\f{\d^2 \psi_k}{\d\cN^2} 
+ \l(\f{1}{H}\,\f{\d H}{\d\cN} - \frac{1}{\cN}\r)\,\f{\d \psi_k}{\d\cN}  
&+& \f{k^2\,\cN^2}{a^2\,H^2}\,\psi_k\nn\\
&=& 2\,\cN\,\f{\d A_k}{\d\cN}  
- \f{20\,a_0^2\,\cN^2\,k_0^2}{a^4\,H^2}\,A_k.\label{eq:eom-psik-cN}
\end{eqnarray}
\end{subequations}
In the above equations, to avoid rather lengthy and cumbersome expressions, we 
have not attempted to express the coefficients involving the conformal time 
coordinate in terms of e-$\cN$-folds.


\subsection{Initial conditions and power spectra}\label{subsec:ic}

Let us now understand the initial conditions that need to be imposed on 
the scalar perturbations.
Note that, at very early times during the contracting phase, the energy 
density of the canonical scalar field $\phi$ dominates the energy density
of the non-canonical field $\chi$.
In inflationary scenarios driven by two fields, it is well known that, when 
the background is largely driven by one of the two fields, the isocurvature 
perturbation can be neglected~\cite{Gordon:2000hv}. 
This seems to suggest that we can ignore the effect of the isocurvature
perturbation on the curvature perturbation at early times.
In such a case, we find that the equation~(\ref{eq:eom-cRk-ss}) governing the 
curvature perturbation simplifies to be
\begin{equation}
\cR_k'+2\,\f{z'}{z}\, \cR_k'+k^2\, \cR_k \simeq 0,\label{eq:eom-cRk-d1}
\end{equation}
where $z\simeq a\, \dot{\phi}/H$, which simplifies to $z\simeq\sqrt{3}\, \Mpl\, 
a$ in the particular matter bounce scenario that we are considering.

\par

In an expanding universe, we can expect the isocurvature perturbations to decay
and, hence, they are not expected to play a significant role at late times.
However, since perturbations can grow in a contracting universe, the effect of 
the isocurvature perturbations may not be negligible as one approaches the
bounce.
Therefore, though the effects of the isocurvature perturbations may be 
insignificant at early times, their contribution may need to be accounted 
for as one approaches the bounce, particularly when the energy density 
of the second field $\chi$ becomes comparable to the energy density of 
the $\phi$ field. 
At early times, we find that the curvature and the isocurvature perturbations 
decouple, and the equation~(\ref{eq:eom-cSk-ss}) describing the isocurvature 
perturbation simplifies to
\begin{equation}
\cS_k''+2\,\f{z'}{z}\, \cS_k'
+\l(\f{k^2}{3}+\f{\phi'^2}{6\, \Mp^2}\r)\,\cS_k\simeq 0.\label{eq:eom-cSk-d1}
\end{equation}

\par

Let us define the Mukhanov-Sasaki variable corresponding to the perturbations
$\cR_k$ and $\cS_k$ to be ${\cal U}_k=z\, \cR_k$ and ${\cal V}_k= z\, \cS_k$,
where $z=a\, \dot{\phi}/H$.
In terms of these variables, in the matter bounce scenario of our interest, the 
above two decoupled equations for $\cR_k$ and $\cS_k$ reduce to
\begin{subequations}
\begin{eqnarray}
{\cal U}_k''+\l(k^2-\f{2}{\eta^2}\r)\, {\cal U}_k &\simeq&0,\label{eq:eom-cUk-d1}\\
{\cal V}_k''+\f{k^2}{3}\, {\cal V}_k &\simeq & 0.\label{eq:eom-cVk-d1}
\end{eqnarray}
\end{subequations}
It is useful to note that, in the matter dominated phase, the mode ${\cal U}_k$ 
behaves exactly as the Mukhanov-Sasaki variable $u_k$ corresponding to the tensor 
perturbation.
At very early times, \ie when $k^2\gg 2/ \eta^2$, we can impose the following 
Bunch-Davies initial conditions on these variables:
\begin{subequations}
\begin{eqnarray}
{\cal U}_k
&=& \f{1}{\sqrt{2\,k}}\, 
{\rm e}^{-i\,k\,\eta},\label{eq:cUk-ic}\\
{\cal V}_k
&=& \f{3^{1/4}}{\sqrt{2\,k}}\,
{\rm e}^{-i\,k\,\eta/\sqrt{3}}.\label{eq:cVk-ic}
\end{eqnarray} 
\end{subequations}
These initial conditions can evidently be translated to the corresponding initial
conditions on $\cR_k$ and $\cS_k$ and their derivatives with respect to the
e-$\cN$-fold.

\par

During early times, when the initial conditions are imposed, the curvature and 
the isocurvature perturbations are considered to be statistically independent 
quantities.
Therefore, as is usually done in the case of two field models, we shall numerically 
integrate the equations~(\ref{eq:eom-cRk-cSk-ss-cN}) using two sets of initial 
conditions (in this context, see, for instance, Refs.~\cite{Tsujikawa:2002qx,
Lalak:2007vi}).
In the first case, we perform the integration by imposing the Bunch-Davies initial 
condition corresponding to~(\ref{eq:cUk-ic}) on $\cR_k$ and setting the initial 
value of $\cS_k$ to be zero.
While, in the second case, we impose the initial condition corresponding 
to~(\ref{eq:cVk-ic}) on $\cS_k$ and set the initial value of $\cR_k$ 
to be zero.
Let us denote the perturbations $\cR_k$ and $\cS_k$ evolved according to these two 
sets of initial conditions to be ($\cR_k^{\rm I}$, $\cS_k^{\rm I}$) and 
($\cR_k^{\rm II}$, $\cS_k^{\rm II}$), respectively.
Then, the power spectra associated with the curvature and the isocurvature 
perturbations can be defined as~\cite{Tsujikawa:2002qx,Lalak:2007vi}
\begin{subequations}
\begin{eqnarray}
\mathcal{P}_{_{\cR}}(k) 
&=& \f{k^3}{2\, \pi^2}\, \l(\l\vert \cR_k^{\rm I} \r\vert^2
+ \l\vert \cR_k^{\rm II} \r\vert^2\r),\\
\mathcal{P}_{_{\cS}}(k)
&=& \f{k^3}{2\, \pi^2}\, \l(\l\vert \cS_k^{\rm I} \r\vert^2
+ \l\vert \cS_k^{\rm II} \r\vert^2\r).
\end{eqnarray}
\end{subequations}


\subsection{Evolution of the perturbations}

\begin{figure}[!t]
\begin{center}
\includegraphics[width=12.00cm]{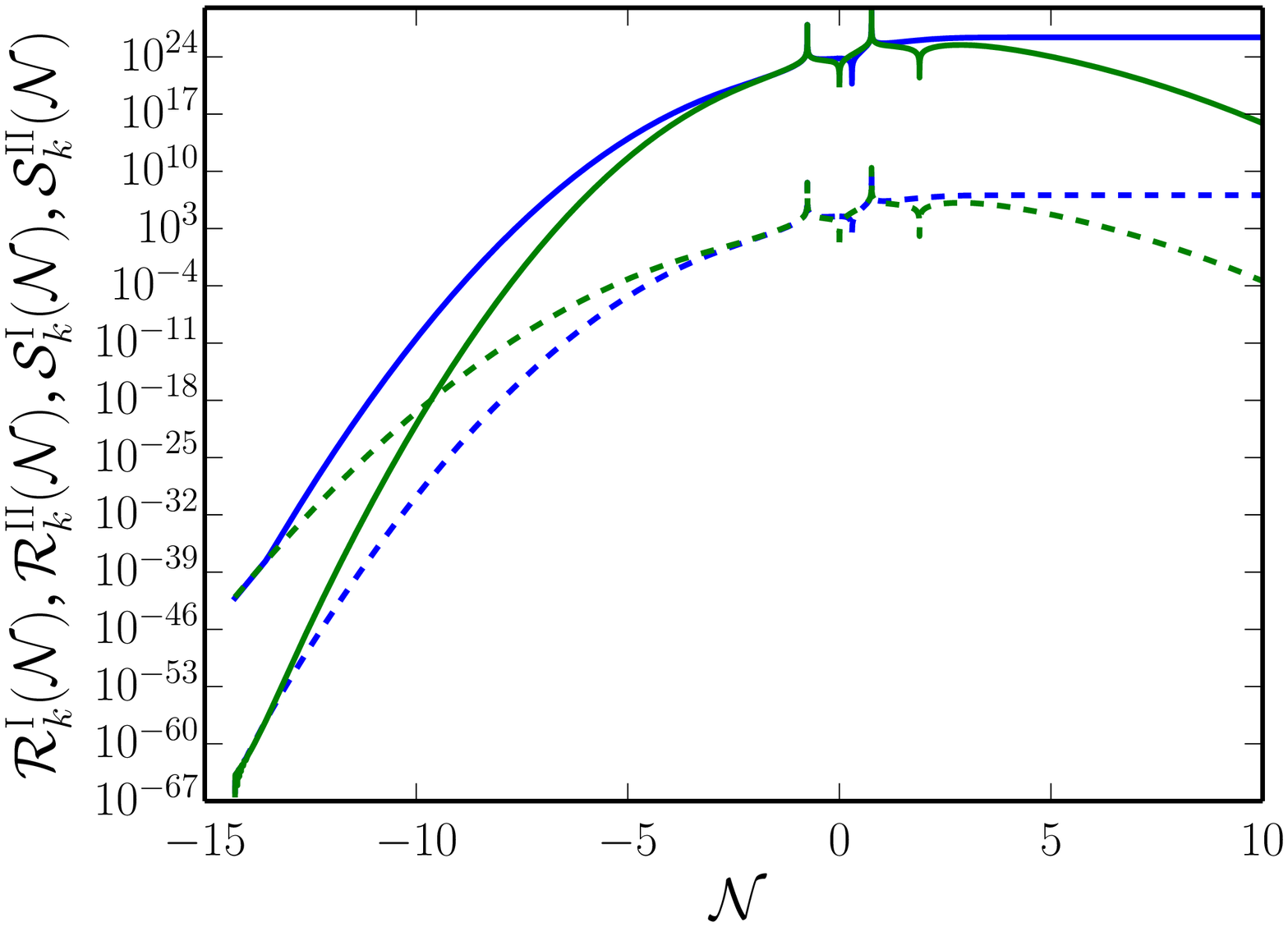} 
\caption{The numerical results for the amplitudes of the curvature (in blue,
with $\cR_k^{\rm I}$ as solid and $\cR_k^{\rm II}$ as dashed) and the isocurvature 
(in green, with $\cS_k^{\rm I}$ as solid and $\cS_k^{\rm II}$ as dashed) 
perturbations evolved with different sets of initial conditions have been plotted 
as a function of e-$\cN$-folds for $k/k_0=10^{-20}$.
As in Fig.~\ref{fig:hk}, we have set $k_0=\Mpl$ and $a_0=3\,\times\,10^7$, 
corresponding to $k_0/(a_0\,\Mpl)=3.3\times 10^{-8}$ which, as we shall see, 
leads to a scale invariant scalar power spectrum whose amplitude matches 
COBE normalization~\cite{Bunn:1996py}.
We have plotted the results from the initial e-$\cN$-fold $\cN_i$ [when $k^2
=10^4\,(a''/a)$] corresponding to the mode.
Note that the amplitudes of $\cR_k^{\rm I}$ and $\cS_k^{\rm I}$ are dominant 
(at suitably late times) when compared to that of $\cR_k^{\rm II}$ and 
$\cS_k^{\rm II}$, respectively.
Also, the curvature perturbation behaves largely in a fashion similar to the tensor 
perturbation [cf.~Fig.~\ref{fig:hk}].
The upward and downward spikes in the plots correspond to points in time where 
the perturbations diverge and vanish, respectively.
As we had expected, both the curvature and the isocurvature perturbations diverge 
at $\eta=\mp\eta_\ast$.
However, it is only the isocurvature perturbation that vanishes at the bounce.
The curvature perturbation actually goes to zero soon after the bounce (during 
$0<\eta<\eta_\ast$), which in turn leads to the vanishing of the isocurvature 
perturbation a little time later (soon after $\eta=\eta_\ast$).
While the curvature perturbation is largely constant (after $\eta=\eta_\ast$)
during the expanding phase, the isocurvature perturbation begins to decay.}
\label{fig:Rk1cSk1cRk2cSk2-n-kbk0-10-20}
\end{center}
\end{figure}

We impose the initial conditions as we have described above when $k^2=10^4\, 
(a^{\prime\prime}/a)$\footnote{Note that, at early times, since $z\propto a$,
$z''/z= a''/a$.
This behavior is indeed expected when the scale factor is described by a power 
law.}.
We integrate the equations~(\ref{eq:eom-cRk-cSk-ss-cN}) governing $\cR_k$ and $\cS_k$ 
from this initial time up to $\eta=-\alpha\, \eta_0$, where, as before, we shall 
set $\alpha=10^5$ (which corresponds to an e-$\cN$-fold of about $\cN \simeq -6.79$).
As in the case of tensors, we carry out the numerical integration using a fifth 
order Runge-Kutta algorithm.
Having integrated for $\cR_k$ and $\cS_k$ until $\cN=-6.79$, we evaluate the 
values of $A_k$ and $\psi_k$ (and their derivatives) at this time by inverting
the relations~(\ref{eq:RS-Apsi}).
Using these as initial conditions, we integrate the equations~(\ref{eq:eom-Ak-psik-cN})
across the bounce until $\eta=\beta\,\eta_0$, with $\beta= 10^2$, which corresponds
to $\cN=4.3$.
We then reconstruct the evolution of $\cR_k$ and $\cS_k$ across the bounce 
using the relations~(\ref{eq:RS-Apsi}).
In Fig.~\ref{fig:Rk1cSk1cRk2cSk2-n-kbk0-10-20}, we have plotted the evolution of 
curvature perturbation $(\cR_k^{\rm I},\cR_k^{\rm II})$ and the isocurvature 
perturbation $(\cS_k^{\rm I}, \cS_k^{\rm II})$, arrived at numerically for a 
specific wavenumber which corresponds to cosmological scales today.

\par

There a few points that needs to be emphasized concerning the results we have
obtained.
As we have discussed earlier, the curvature and the isocurvature perturbations
are expected to diverge at $\eta=\mp\eta_\ast$, and it is clear from the figure
that they indeed do so.
Moreover, the isocurvature perturbation vanishes at the bounce, as expected.
In contrast, we find that the curvature perturbation does not vanish at the
bounce as one may naively guess, but does so a little time after the bounce.
This behavior seems to be responsible for the isocurvature perturbation too
to vanish a little time later.
Note that, as in the case of tensors, the amplitude of the curvature perturbation 
almost freezes at suitably late times (in fact, after $\eta=\eta_\ast$) during 
the expanding phase.
During the period, the isocurvature perturbations begin to decay.
As we shall illustrate later, this leads to a strongly adiabatic scalar power
spectrum, with the amplitude of the isocurvature perturbations being much
smaller than the curvature perturbations.

\par


\section{Analytical arguments}\label{sec:a}

In this section, we shall arrive at analytical solutions for the curvature 
and isocurvature perturbations for scales of cosmological interest under 
well-motivated approximations.
We shall again divide the period of interest into two domains, as we have 
discussed already.
Let us go on to construct the solutions to the equations governing the scalar 
perturbations in the two domains.


\subsection{Solutions in the first domain}

As we have discussed before, at early times during the matter dominated 
contraction, we can assume that the equations governing the evolution 
of the curvature and the isocurvature perturbations are decoupled.
We had mentioned earlier that, during this phase, the mode ${\cal U}_k$
is expected to behave exactly like the Mukhanov-Sasaki variable $u_k$
associated with the tensor mode.
This is not surprising since such a behavior is well known in power law
expansion and, hence, can be expected in power law contraction as well.
Using the Bunch-Davies initial condition~(\ref{eq:cUk-ic}), during 
sufficiently early times, the solution to Eq.~(\ref{eq:eom-cRk-d1}) can be 
obtained to be
\begin{equation}
\cR_k(\eta)
\simeq \f{1}{\sqrt{6\, k}\,\Mp\,a_0\, k_0^2 \,\eta^2}\,
\l(1-\f{i}{k\, \eta}\r)\, {\rm e}^{-i\,k\,\eta},\label{eq:cRk-d1}
\end{equation}
which, it should be emphasized, is the same as the solution~(\ref{eq:hk-d1})
for the tensor mode apart from an overall constant.

\par

Obtaining the solution to the isocurvature perturbation requires a little
more care.
In arriving at the equation~(\ref{eq:eom-cVk-d1}) governing the Mukhanov-Sasaki
variable associated with the isocurvature perturbation, we had completely 
ignored the role of the curvature perturbation.
While this seems acceptable for determining the initial condition, we find that 
the effect of the curvature perturbation needs to be accounted for, in order
to achieve a better approximation.
During the first domain, upon using the expression~(\ref{eq:cRk-d1}) for 
$\cR_k$, we find that the solution to Eq.~(\ref{eq:eom-cSk-ss}) can be 
obtained to be~\cite{Mathematica8.0}
\begin{eqnarray}
\cS_k(\eta) 
&\simeq& \f{1}{9\,\sqrt{2\,k^3}\,a_0\, k_0^3\, \Mp \eta^4}\,
\biggl(-12\,i\,\l(1+i\,k\,\eta\r)\,{\rm e}^{-i\,k\,\eta}
+ \f{9}{3^{1/4}}\,k\,k_0\,\eta^2\,{\rm e}^{-i\,k\,\eta/\sqrt{3}}\nn\\
&+& 4\,k^2\,\eta^2\,{\rm e}^{-i\,k\,\eta/\sqrt{3}}\, \l\{\pi+i\,
{\rm Ei}\l[{\rm e}^{-i\,(3-\sqrt{3})\,k\,\eta/3}\r]\r\}\biggr),\label{eq:cSk-d1}
\end{eqnarray}
where ${\rm Ei}\l[z\r]$ is the exponential integral function (see, for
instance, Ref.~\cite{Gradshteyn:2007}).
It is straightforward to check that, at early times, it is the second term
in the above expression which survives, which exactly corresponds to the
initial condition~(\ref{eq:cVk-ic}).


\subsection{Solutions in the second domain}

As we have discussed, in the second domain (\ie over the period $-\alpha\,
\eta_0 \le \eta \le \beta\,\eta_0$), we shall solve for the metric perturbations 
$A_k$ and $\psi_k$.
In this domain, for scales of cosmological interest, we can ignore the $k$ 
dependent terms in Eqs.~(\ref{eq:eom-Ak-psik-eta-ss}). 
Under this condition, the two equations can be combined to obtain that
\begin{equation}
(A_k+\psi_k)'' + 2\,\mathcal{H}\,(A_k+\psi_k)'\simeq 0,
\end{equation}
which is exactly the equation for the tensor mode $h_k$ that we had arrived
in this domain [cf.~Eq.~(\ref{eq:eom-hk-d2})].
This equation can be integrated once to yield 
\begin{equation}
(A_k+\psi_k)' \simeq \f{k_0\,\cC_k}{a^2}.
\end{equation}
with $\cC_k$ being a constant of integration. 
Upon further integration, we obtain that
\begin{equation}\label{eq:Apluspsi}
A_k(\eta)+\psi_k(\eta) 
\simeq \f{\cC_k}{2\,a_0^2}\,f(k_0\,\eta) + \cD_k,
\end{equation}
where the function $f$ is given by Eq.~(\ref{eq:f}) and $\cD_k$ is 
a second constant of integration. 
Upon substituting this result in Eq.~(\ref{eq:eom-Ak-eta-ss}), we can arrive 
at an equation governing $A_k$.
On solving the resulting differential equation (say, using 
Mathematica~\cite{Mathematica8.0}), we find that the solution for $A_k$ is 
given by
\begin{equation}
A_k(\eta)
\simeq \f{\cC_k\, k_0\,\eta}{4\, a_0^2\, (1+k_0^2\,\eta^2)}
+\cE_k\, {\rm e}^{-2\,\sqrt{5}\,\tan^{-1}(k_0\, \eta)}
+ \cF_k\, {\rm e}^{2\,\sqrt{5}\,\tan^{-1}(k_0\, \eta)},
\end{equation}
where $\cE_k$ and $\cF_k$ denote two additional constants of integration.
The corresponding solution for $\psi_k$ can be obtained by substituting this
result in Eq.~(\ref{eq:Apluspsi}). 

\par

Having obtained the solutions for $A_k$ and $\psi_k$, we can now reconstruct 
the curvature and the isocurvature perturbations $\cR_k$ and $\cS_k$ using
Eqs.~(\ref{eq:RS-Apsi}). 
We find that, in the second domain, $\cR_k$ and $\cS_k$ are given by
\begin{subequations}
\begin{eqnarray}\label{eq:R-a1}
\cR_k(\eta) 
&\simeq& 
\f{-1}{2\,a_0^2\,\l(1 - 2\,k_0^2\,\eta^2 - 3\,k_0^4\,\eta^4\r)}\,
\Biggl(\cC_k\,\biggl[\l(1 + 3\,k_0^2\,\eta^2\r)\, k_0\,\eta\nn\\
& &-\, \l(1 - 2\,k_0^2\,\eta^2 - 3\,k_0^4\,\eta^4\r)\,
\tan^{-1}\l(k_0\,\eta\r)\biggr]\nn\\
& &-\, 2\,a_0^2\,\l(1 + k_0^2\,\eta^2\r)\,
\biggl[\cD_k\,\l(1 - 3\,k_0^2\,\eta^2\r)
-\cE_k\,\l(1 + 2\,\sqrt{5}\,k_0\,\eta - k_0^2\,\eta^2\r)\,
{\rm e}^{-2\,\sqrt{5}\tan^{-1}(k_0\,\eta)}\nn\\
& &-\, \cF_k\,\l(1 - 2\,\sqrt{5}\,k_0\,\eta - k_0^2\,\eta^2\r)\,
{\rm e}^{2\,\sqrt{5}\tan^{-1}(k_0\,\eta)}\biggr]\biggr\},
\label{eq:cRk-d2}\\
\cS_k(\eta) 
&\simeq& \frac{-k_0\,\eta}{2\,\sqrt{3}\,a_0^2\,
\l(1 + k_0^2\,\eta^2\r)^{1/2}\,\l(1 - 3\,k_0^2\,\eta^2\r)}\,
\Biggl\{3\,\cC_k+ 8\,a_0^2\,\biggl[\cE_k\,\l(\sqrt{5} + k_0\,\eta\r)\,
{\rm e}^{-2\,\sqrt{5}\,\tan^{-1}(k_0\,\eta)}\nn\\
& &-\,\cF_k\,\l(\sqrt{5} - k_0\,\eta\r)
{\rm e}^{2\,\sqrt{5}\,\tan^{-1}(k_0\,\eta)}\biggr]\Biggr\}.\label{eq:cSk-d2}
\end{eqnarray}
\end{subequations}
The four constants $\cC_k$, $\cD_k$, $\cE_k$ and $\cF_k$ can be determined by 
matching these solutions with the solutions for $\cR_k$ and $\cS_k$ we had
obtained in the first domain at $\eta=-\alpha\,\eta_0$.
The expressions describing the constants are long and cumbersome and, hence,
we relegate the details to an appendix (see App.~\ref{appendix:1}).


\subsection{Comparison with the numerical results}

\begin{figure}[!htb]
\begin{center}
\includegraphics[width=12.00cm]{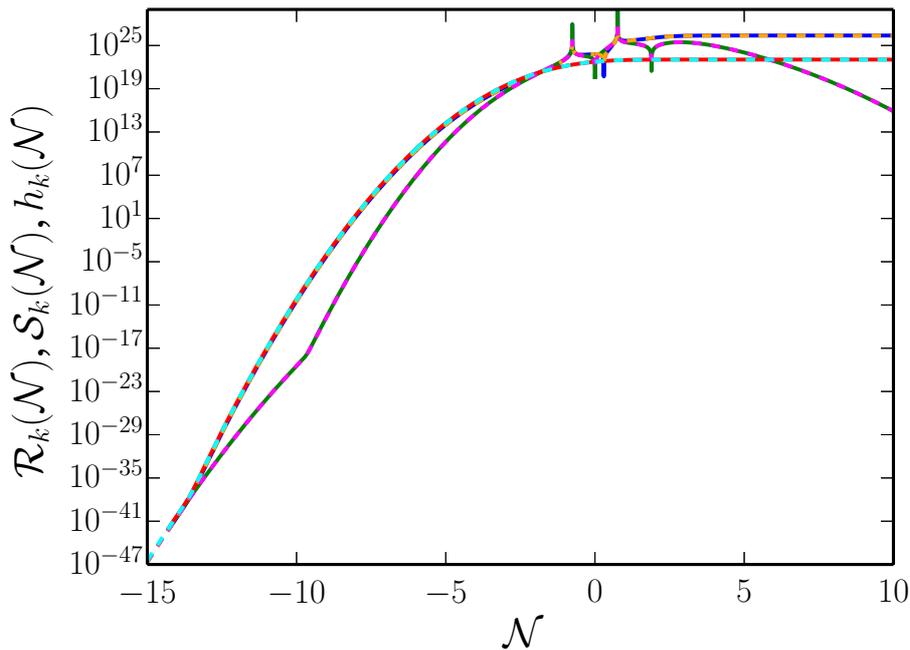}
\caption{A comparison of the numerical results (solid lines) with the analytical 
results (dashed lines) for the amplitude of the curvature perturbation $\cR_k$ 
(blue solid line and orange dashed line), the  isocurvature perturbation $\cS_k$ 
(green solid line and magenta dashed line) and the tensor mode $h_k$ (red solid 
line and cyan dashed line) corresponding to the wavenumber $k/k_0 
= 10^{-20}$.  
As earlier, we have set $k_0=\Mpl$ and $a_0=3\,\times\,10^7$, 
corresponding to $k_0/(a_0\,\Mpl)=3.3\times 10^{-8}$ and, for plotting 
the analytical results, we have chosen $\alpha=10^5$.
We have plotted the numerical results from the initial e-$\cN$-fold $\cN_i$ 
[when $k^2=10^4\,(a''/a)$] corresponding to the mode.
Evidently, the analytical and numerical results match extremely well, suggesting 
that the analytical approximation for the modes works to a very good accuracy.
Notice that, around the bounce, the amplitude of the scalar perturbations are 
enhanced by a few orders of magnitude more than that of the tensor perturbations.
It is this feature, which is obviously a result of the specific behavior of the 
background near the bounce, that leads to a viable tensor-to-scalar ratio.}
\label{fig:cRkcSkhk-comp-kbk0-10-20}
\end{center}
\end{figure}

\begin{figure}[!htb]
\begin{center}
\includegraphics[width=12.00cm]{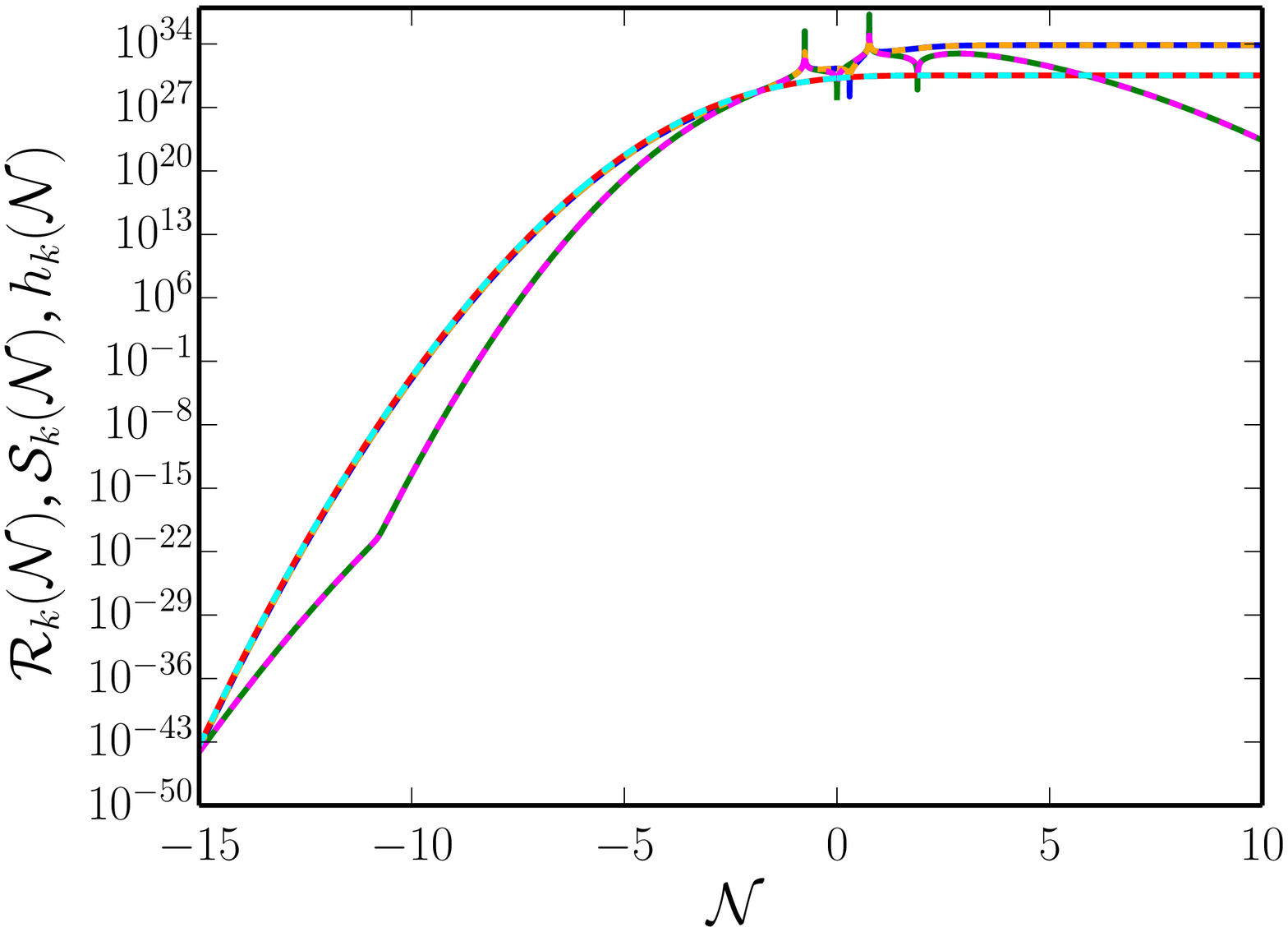} 
\caption{The plots as in the previous figure for the wavenumber $k/k_0 
= 10^{-25}$.
Clearly, the analytical results are in good agreement with the numerical 
results.}
\label{fig:cRkcSkhk-comp-kbk0-10-25}
\end{center}
\end{figure}

Let us now compare the above analytical results for $\cR_k$ and $\cS_k$ with 
the numerical results. 
Recall that, numerically, we had obtained two sets of solutions for $\cR_k$ 
and $\cS_k$, \viz $(\cR_k^{\rm I},\cS_k^{\rm I})$ and 
$(\cR_k^{\rm II},\cS_k^{\rm II})$, corresponding to two different sets of 
initial conditions.
In contrast, while arriving at the analytical results, for convenience, we have 
imposed the Bunch-Davies initial on both $\cR_k$ and $\cS_k$ simultaneously.
We shall compare the amplitudes of $\cR_k$ and $\cS_k$ obtained analytically 
with the amplitudes $\cR_k^{\rm I}+\cR_k^{\rm II}$ and $\cS_k^{\rm I}+\cS_k^{\rm II}$
arrived at numerically.
(Recall that, the amplitudes of $\cR_k^{\rm I}$ and $\cS_k^{\rm I}$ had dominated
those of $\cR_k^{\rm II}$ and $\cS_k^{\rm II}$, respectively.)
In Figs.~\ref{fig:cRkcSkhk-comp-kbk0-10-20} and~\ref{fig:cRkcSkhk-comp-kbk0-10-25}, 
we have plotted the analytical and the numerical results for wavenumbers such that 
$k/k_0 = 10^{-20}$ and  $k/k_0 = 10^{-25}$, respectively.
As is evident from the figures, the analytical results match the numerical results 
very well.
In fact, we find the difference between the analytical and numerical results 
to be less than $2\%$.


\section{The scalar power spectra and 
the tensor-to-scalar ratio}\label{sec:sps-r}

With the analytical and the numerical results at hand, let us now go on to 
evaluate the scalar power spectra and the tensor-to-scalar ratio.
In order to understand the effects of the bounce on these quantities, let 
us evaluate the scalar and tensor power spectra before as well as after the
bounce.

\par

\begin{figure}[!htb]
\begin{center}
\includegraphics[width=12.00cm]{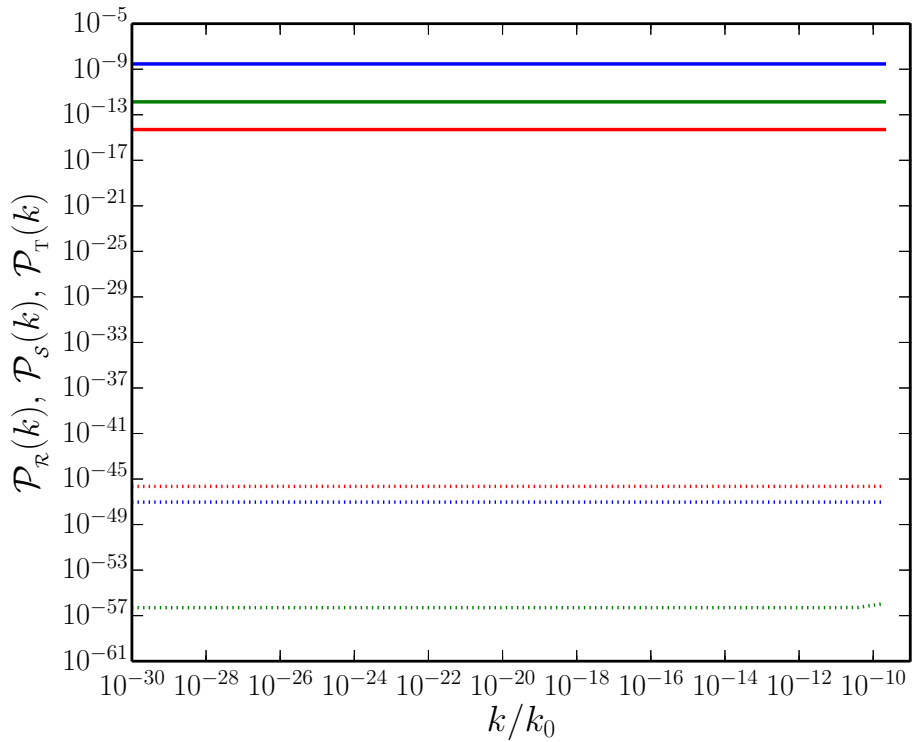} 
\caption{The numerically evaluated scalar (the curvature perturbation 
spectrum in blue and the isocurvature perturbation spectrum in green) 
and tensor power spectra (in red) have been plotted as a function of 
$k/k_0$ for a wide range of wavenumbers. 
The power spectra have been plotted both before the bounce (as dotted 
lines) and after (as solid lines).
The power spectra have been evaluated at $\eta=-\alpha\, \eta_0$ (with
$\alpha=10^5$) before the bounce and at $\eta=\beta\,\eta_0$ (with $\beta
=10^2$) after the bounce.
In plotting the figure, we have set $k_0/(a_0\,\Mpl)=3.3\,\times\,10^{-8}$, 
as in the previous figures.
All the power spectra are evidently scale invariant over scales of 
cosmological interest.
Also, the above choice of $k_0/a_0$ leads to a COBE normalized curvature
perturbation spectrum.
Moreover, the tensor-to-scalar ratio evaluated after the bounce proves to 
be rather small ($r\simeq 10^{-6}$), which is consistent with the current 
upper limits on the quantity.}
\label{fig:ps-n}
\end{center}
\end{figure}

Let us first consider the numerical results, which are exhibited in 
Fig.~\ref{fig:ps-n}.
All the power spectra are strictly scale invariant (over scales of cosmological
interest) before as well as after the bounce.
The power spectra before the bounce have been evaluated at $\eta=-\alpha\,\eta_0$,
with $\alpha=10^5$, which, as we had mentioned, corresponds to $\cN=-6.79$.
The power spectra after the bounce have been evaluated at $\eta=\beta\, \eta_0$,
with $\beta=10^2$, which, recall that, corresponds to $\cN=4.3$.
Since the scales of cosmological interest are much smaller than the 
scale associated with the bounce, the shapes of the power spectra are 
indeed expected to remain unaffected by the bounce.
While a bounce generically enhances the amplitude of the perturbations,
the scalar and tensor perturbations can be expected to be amplified by 
different amounts, depending on the behavior of the background close
to the bounce.
Note that, in the scenario of our interest, the tensor-to-scalar ratio 
is rather large before the bounce.
In fact, the tensor-to-scalar ratio well before the bounce proves to be 
of the order of ${\cal O}(24)$, a result that is well known in the 
literature (see, for instance, Ref.~\cite{Allen:2004vz}).
As we had pointed out, in our case, the bounce amplifies the scalar 
perturbations much more than the tensor perturbations
[cf. Figs.~\ref{fig:cRkcSkhk-comp-kbk0-10-20} 
and~\ref{fig:cRkcSkhk-comp-kbk0-10-25}].
In other words, the bounce suppresses the tensor-to-scalar ratio.
Recall that, the only parameter that occurs in our model is the 
combination $k_0/a_0$.
We find that, for a choice of $k_0/a_0$ that leads to a COBE normalized
scalar power spectrum after the bounce, \ie ${\cal P}_{_{\cR}}(k)
\simeq 2.31\times 10^{-9}$~\cite{Bunn:1996py}, the corresponding 
tensor-to-scalar ratio proves to be much smaller than the current 
upper bound of $r<0.1$ from Planck~\cite{Ade:2015lrj}.
It is also useful to note that isocurvature perturbations, while they
grow across the bounce, begin to decay at late times (actually, after
$\eta>\eta_\ast$).
At a sufficiently late time when we evaluate the power spectra, their
amplitude proves to be about four orders of magnitude smaller than the
amplitude of the curvature perturbation.
This suggests that the power spectrum is strongly adiabatic, which is
also consistent with the recent observations~\cite{Ade:2015lrj}.

\par

\begin{figure}[!htb]
\begin{center}
\includegraphics[width=12.00cm]{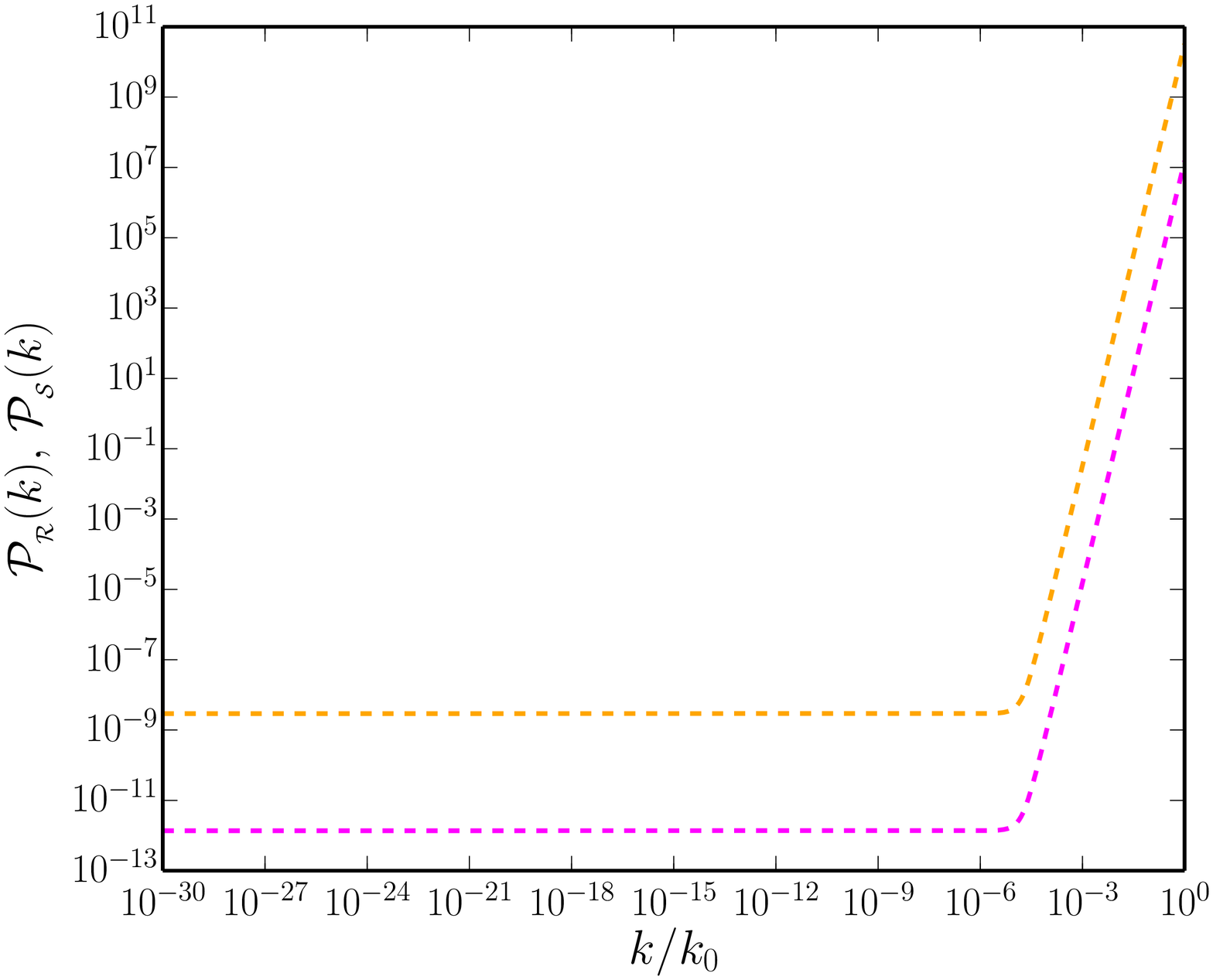} 
\caption{The curvature (in orange) and the isocurvature (in magenta) 
perturbation spectra evaluated analytically after the bounce.
In plotting this figure, we have chosen the same values for the 
various parameters as in Fig.~\ref{fig:tps-a}, wherein we had
plotted the tensor power spectrum obtained analytically.
As in the case of the tensor power spectrum, these analytical spectra 
are valid only for $k\ll k_0/\alpha$.
We find that the scale invariant amplitudes at such small wavenumbers
match the numerical results presented in the previous figure very well.}
\label{fig:sps-a}
\end{center}
\end{figure}

Let us now evaluate the scalar power spectra analytically after the bounce.
At a sufficiently late time after the bounce (say, when $\eta\gg\eta_\ast$), 
we find that the curvature perturbation turns almost a constant
[cf. Eq.~(\ref{eq:cRk-d2})], and is given by
\begin{equation}
\cR_k(\eta)
\simeq \cC_k\, \f{\pi}{4\,a_0^2} - \cE_k\, \f{{\rm e}^{-\sqrt{5}\pi}}{3} 
- \cF_k\, \f{{\rm e}^{\sqrt{5}\,\pi}}{3} + \cD_k.
\end{equation}
We have plotted the power spectrum associated with this curvature perturbation
in Fig.~\ref{fig:sps-a}, which is very similar in shape to the analytical 
tensor power spectrum we had plotted earlier [cf.~Fig.~\ref{fig:tps-a}].
Note that our analytical approximations are valid only when $k\ll k_0/\alpha$,
and the spectrum is indeed scale invariant over this domain, reflecting the
behavior obtained numerically.
If we now assume that $k\ll k_0/\alpha$, we obtain the scale invariant
amplitude of the curvature perturbation spectrum to be
\begin{equation}
\mathcal{P}_{_{\cR}}(k)
\simeq \frac{k_0^2\,{\rm e}^{4\,\sqrt{5}\,\pi}}{61440\,\pi^2\,a_0^2\,\Mp^2},
\label{eq:sps-sia}
\end{equation}
which we find matches the numerical result [of COBE normalized amplitude for
$k_0/(a_0\,\Mpl)=3.3\times10^{-8}$] very well.

\par

\begin{figure}[!htb]
\begin{center}
\includegraphics[width=12.00cm]{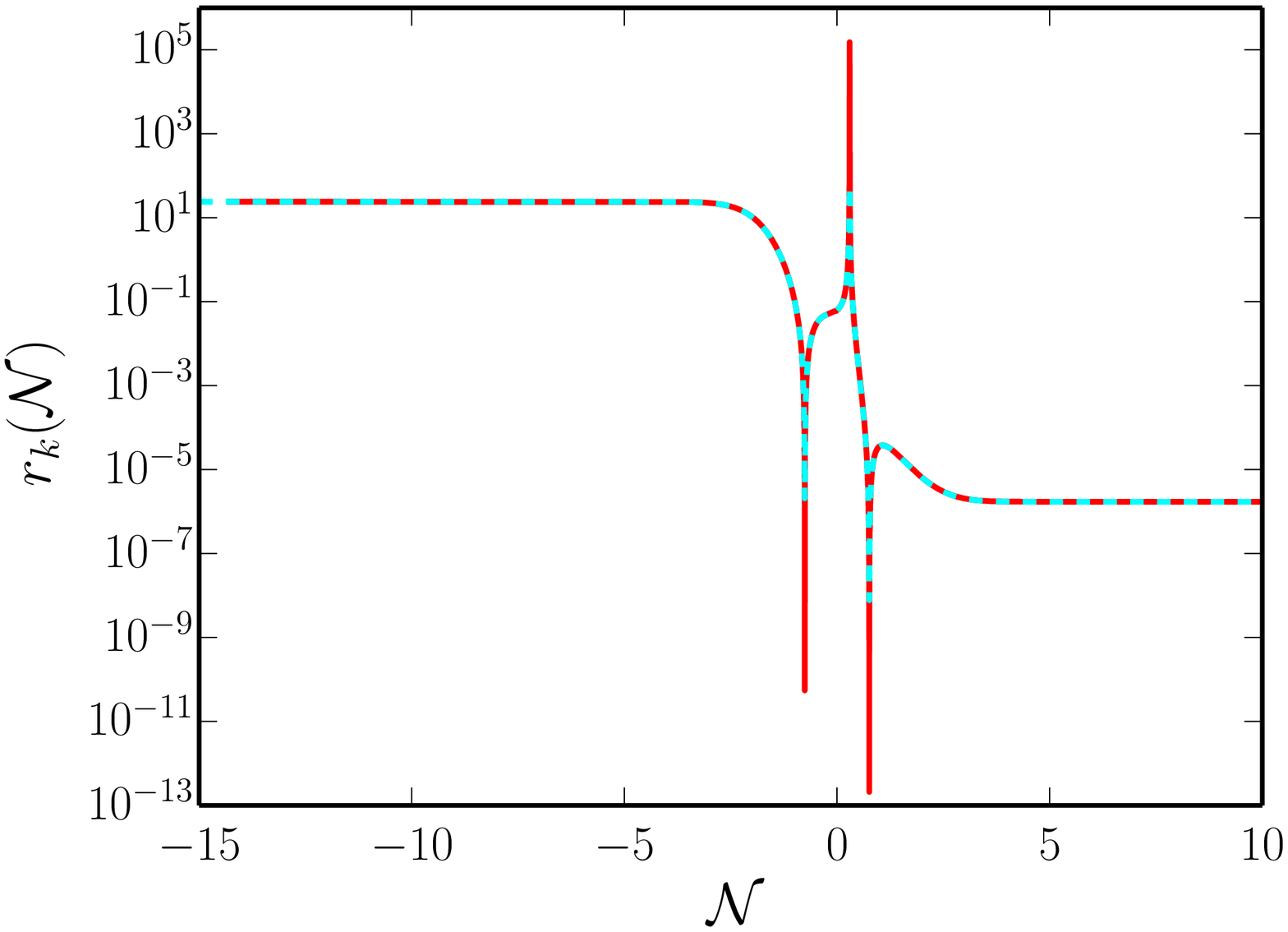} 
\caption{The tensor-to-scalar ratio calculated numerically (red solid line) 
and analytically (cyan dashed line) have been plotted as a function of 
$\mathcal{N}$ for the wavenumber $k/k_0=10^{-20}$.
The numerical and the analytical results agree well as expected.
Note that the bounce suppresses the tensor-to-scalar ratio from a large value 
($r_k \simeq 20$) to a rather small value ($r_k\simeq 10^{-6}$).}
\label{fig:r-na} 
\end{center}
\end{figure}

From the analytical and numerical results for the scalar and tensor modes, 
we can also understand the behavior of the tensor-to-scalar ratio across 
the bounce.
In Fig.~\ref{fig:r-na}, we have plotted the evolution of the 
tensor-to-scalar $r_k=\mathcal{P}_{_{\rm T}}(k)/\mathcal{P}_{_{\cR}}(k)$ 
for a given mode with wavenumber $k/k_0=10^{-20}$.
We have plotted the numerical as well as the analytical results in the figure.
The numerical and the analytical results agree well with each other.
Also, $r_k$ vanishes (at $\eta=\mp\eta_\ast$) and diverges (during
$0<\eta<\eta_\ast$) exactly reflecting the behavior of the curvature 
perturbation (which diverges and vanishes at these points, respectively).
Importantly, the bounce suppresses the tensor-to-scalar ratio from a large 
value ($r_k \simeq 20$) to a rather small value ($r_k\simeq 10^{-6}$) that
is consistent with the current upper bounds.
It is interesting to note that tensor-to-scalar ratio is a pure number and
is actually independent of even the single parameter $k_0/a_0$ that 
characterizes our model [cf.~Eqs.~(\ref{eq:tps-sia}) and~(\ref{eq:sps-sia})].

\par

Our last task is to arrive at the isocurvature power spectrum analytically. 
At large times after the bounce (such that $\eta \gg\eta_\ast)$, the behavior
of the isocurvature perturbation can be shown to be [cf. Eq.~(\ref{eq:cSk-d2})]
\begin{equation}
\cS_k(\eta) \simeq \frac{4\,\cF_k\,{\rm e}^{\sqrt{5}\,\pi}}{3\,\sqrt{3}\,k_0\,\eta}.
\end{equation}
Unlike the curvature perturbation, the isocurvature perturbation is not a constant 
in this domain, but decays with the expansion of the universe. 
This behavior is also evident from the numerical results [cf. 
Figs.~\ref{fig:cRkcSkhk-comp-kbk0-10-20} and~\ref{fig:cRkcSkhk-comp-kbk0-10-25}].
For scales of cosmological interest such that $k\ll k_0/\alpha$, we find that
the isocurvature perturbation spectrum, evaluated at $\eta=\beta\,\eta_0$, is
given by
\begin{equation}
\mathcal{P}_{_{\cS}}(k) 
\simeq \frac{k_0^2\,{\rm e}^{4\,\sqrt{5}\,\pi}}{11520\,\beta^2\,\pi^2\,a_0^2\,\Mp^2}.
\label{eq:ica}
\end{equation}
For the values of the parameters we have been working with, \viz $k_0/(a_0\,\Mpl)
=3.3\times 10^{-8}$ and $\beta=10^2$, we find that the above analytical estimate 
agrees well with the numerical results we have obtained.


\section{Summary and outlook}\label{sec:so}

One of the problems that had plagued completely symmetric bouncing scenarios 
is the fact that the tensor-to-scalar ratio in such models proves to be 
large, typically well beyond the current constraints from the cosmological
data.
In this work, we have constructed a two field model consisting of a canonical
scalar field and a non-canonical ghost field to drive a symmetric matter bounce
and have studied the evolution of the scalar and tensor perturbations in the 
model.
For a specific choice of the scale factor describing the matter bounce, we have 
been able to arrive at completely analytical solutions for all the background 
quantities.
We find that the model we have constructed involves only one parameter, \viz
the ratio of the scale associated with the bounce to the value of the scale 
factor at the bounce.
Using the background solutions, we have numerically evolved the perturbations
across the bounce and have evaluated the scalar and tensor power spectra
after the bounce.
In order to circumvent the issues confronting the evolution of the curvature
and the isocurvature perturbations in a bouncing scenario, we have worked in
a specific gauge wherein the two independent scalar perturbations behave well
across the bounce.
Once having evolved the perturbations, we reconstruct the curvature and the 
isocurvature perturbations from these quantities and evaluate the corresponding
power spectra.
We show that the scalar and tensor perturbation spectra in our model prove to 
be strictly scale invariant, as is expected to occur in a matter bounce
scenario.
We also explicitly illustrate a well understood result, \viz while the 
bounce affects the amplitudes of the power spectra, their shapes remain 
unmodified across the bounce over scales of cosmological interest. 
Moreover, we find that, for a value of the scale factor that leads to the COBE 
normalized power spectrum for the curvature perturbation, the tensor-to-scalar
ratio proves to be of the order of $r\simeq 10^{-6}$, which is, obviously,
perfectly consistent with the current upper bounds from the recent CMB 
observations. 
Further, we have shown that, the amplitude of the isocurvature perturbations are
quite small (their power spectrum is about four orders of magnitude below the
power spectrum of the curvature perturbation).
This indicates that the scenario generates a strongly adiabatic scalar perturbation 
spectra, again an aspect which is consistent with the observations.  
Importantly, we also support all the numerical results with analytical arguments.

\par

Before, we conclude, we believe we should clarify a few different points. 
As we have pointed out repeatedly, our model essentially
depends on only one parameter, \viz\/ $k_0/a_0$.
This is evident from potential governing the model and this aspect 
is also reflected in the results we have obtained.
Notice that the amplitudes of the scalar as well as the tensor power spectra [cf. 
Eqs.~(\ref{eq:tps-sia}), (\ref{eq:sps-sia}) and~(\ref{eq:ica})] actually
depend only on the ratio $k_0/(\Mpl\, a_0)$.
We have chosen $k_0/(a_0\,\Mpl)= 3.3\times 10^{-8}$ in order to lead to a COBE 
normalized curvature perturbation spectrum. 
It is also important to note that, in terms of cosmic time, the duration of 
the bounce is, in fact, of the order of $a_0\,\eta_\ast\simeq a_0/k_0$.

\par

Another point that requires some clarification is the assumption of a 
symmetric bounce.
The assumption of a symmetric bounce has proved to be convenient for us 
to study the problem. 
Actually, the scale factor and the model that drives the background will 
be valid only until an early epoch after the bounce. 
Hence, the term symmetric bounce basically refers to the period close to 
the bounce. 
At a suitable time after the bounce, we expect the energy from the scalar 
fields to be transferred to radiation as is done, for instance, in 
perturbative reheating after inflation. 
Though we have not touched upon this issue here, we believe that reheating 
can be achieved with a simple coupling (such as the conventional $\Gamma\, 
\dot{\phi}$ term) between the scalar field and the radiation fluid. 
Since we expect reheating to be achieved in such a fashion, we have ignored 
the presence of a radiation fluid in this work.

\par

While it is interesting to have achieved a tensor-to-scalar ratio that is 
consistent with the observations in a completely symmetric matter bounce
scenario, needless to add, many challenges remain.
Theoretically, the model needs to be examined in greater detail to understand
the fundamental reason as to why it leads to a small tensor-to-scalar ratio.
In this context, the best way forward seems to be to consider different models 
leading to the same factor and investigate the behavior of the perturbations 
in these different models.
Another related point is regarding the concern that has been raised about the 
situations under which the standard initial conditions can be imposed (in this
context, see, for instance, Ref.~\cite{Peter:2015zaa}).
In the case of our model, since the curvature and the isocurvature perturbations
decouple during the early contracting phase, we have been able to impose the 
standard Bunch-Davies initial conditions.

From an observational point of view, we need to generate a tilt in the scalar 
power spectrum to match the CMB observations.
Moreover, we need to examine if the scalar non-Gaussianities generated in the 
model are indeed consistent with the current constraints from 
Planck~\cite{Ade:2015ava}. 
Further, rather than brush them aside, we need to get around to addressing the 
different theoretical issues plaguing bouncing models that we had discussed in 
some detail in the introductory section.
We should point out here that a completely nonperturbative analysis of a model
very similar to what we have considered seems to suggest such models may not be 
as pathological as it has been argued to be~\cite{Xue:2013bva}.  
Clearly, one needs to explore more complex models beyond the simple model we 
have constructed here.
We are presently investigating a variety of such issues. 


\section*{Acknowledgements}

DC would like to thank the Indian Institute of Technology Madras, Chennai, 
India, for financial support through half-time research assistantship.
LS wishes to thank Rishi Khatri for discussions.
The authors thank Robert Brandenberger, Patrick Peter and V.~Sreenath for 
comments on the manuscript.


\appendix

\section{Fixing the coefficients}\label{appendix:1}
 
Recall that, in Sec.~\ref{sec:a}, the expressions~(\ref{eq:cRk-d2}) 
and~(\ref{eq:cSk-d2}) that describe the analytical solutions for 
$\cR_k$ and $\cS_k$ in the second domain had contained four 
time-independent constants, \viz $\cC_k$, $\cD_k$, $\cE_k$ and $\cF_k$.
As we had described, these four constants are to be determined by
matching the solutions for $\cR_k$ and $\cS_k$ in the first domain 
[cf. Eqs.~(\ref{eq:cRk-d1}) and (\ref{eq:cSk-d1})] and their time 
derivatives with the corresponding quantities in the second domain.
This matching has to be carried out at the junction of the two
domains, \viz at $\eta=-\alpha\,\eta_0$.
These matching conditions lead to four equations which need to be 
solved simultaneously to arrive at the constants.
The constants can be determined to be
\begin{eqnarray}
\cC_k 
&=& \f{\l(\alpha^2 + 1\right)\,a_0}{54\,\sqrt{2}\,\alpha^7\,k_0^2\,\Mp\,k^{3/2}\,}\,
\Biggl\{16\,\alpha^2\,\sqrt{\alpha^2 + 1}\;k^2\nn\\
& &\times\,\l[\alpha\,\l(\alpha^2 + 1\r)\,k
- 3\,\sqrt{3}\,i\,\l(\alpha^2 - 1\r)\,k_0\r]\,
{\rm Ei}\l[i\,\l(3 - \sqrt{3}\r)\,\alpha\,k/(3\,k_0)\r]\,
{\rm e}^{i\,\alpha\,k/(\sqrt{3}\,k_0)}\nn\\
& &+\, \sqrt{3}\,\Biggl[i\,\alpha^2\,\l(\alpha^2 + 1\r)\,
\l(-9\,\alpha^3+ 32\,\sqrt{\alpha^2 + 1} + 27\,\alpha\r)\,k_0\,k^2\nn\\ 
& &+\,\l(9\,\alpha^5 - 18\,\alpha^3 
+ 16\,\alpha^2\,\sqrt{\alpha^2 + 1} 
- 80\,\sqrt{\alpha^2 + 1} - 27\,\alpha\r)\,\l(3\,\alpha\,k_0^2\,k+3\,i\,k_0^3\r)\Biggr]\,
{\rm e}^{i\,\alpha\,k/k_0}\,\nn\\ 
& &-\, 4\,\alpha^2\,\sqrt{\alpha^2 + 1}\,
\l(4\,\pi\,k^2 + 3^{7/4}\,k_0\,k\r)\,\!\!
\l[3\,\sqrt{3}\,\l(\alpha^2 - 1\r)\,k_0 
+ i\,\alpha\,\l(\alpha^2 + 1\right)\,k\r]\,
{\rm e}^{i\,\alpha\,k/(\sqrt{3}\,k_0)}\Biggr\},\nn\\
\end{eqnarray}
\begin{eqnarray}
\cD_k 
&=& \frac{1}{108\,\sqrt{2}\;3^{3/4}\,\alpha^7\,a_0\,k_0^2\,\Mp\,k^{3/2}}\,
\Biggl\{4\;3^{1/4}\,\alpha^2\,\sqrt{\alpha^2 + 1}\,
\biggl(\sqrt{3}\,\alpha^2\,\l(\alpha^2 + 1\r)\,k^3\nn\\ 
& &-\, 9\,i\,\alpha\,\l(4\,\alpha^2 - 1\r)\,k_0\,k^2
+ \l[4\,\sqrt{3}\,\alpha\,\l(\alpha^2 + 1\r)^2\,k 
- 36\,i\,\l(\alpha^4 - 1\r)\,k_0\r]\,k^2\,\tan^{-1}(\alpha)\biggr)\nn\\
& &\times\,{\rm Ei}\l[i\,\l(3 - \sqrt{3}\r)\,\alpha\,k/(3\,k_0)\r]\,
{\rm e}^{i\,\alpha\,k/(\sqrt{3}\,k_0)}\nn\\\nn\\ 
& &+\, 3^{5/4}\,\alpha\, \biggl[i\,\alpha^2\,\l(\alpha^2 + 1\r)\,
\l(-9\,\alpha^3 + 8\,\sqrt{\alpha^2 + 1} + 9\,\alpha\r)\,k_0\,k^2 \nn\\
& &+\,\l(9\,\alpha^5 + 12\,\alpha^3 + 40\,\alpha^2\,\sqrt{\alpha^2 + 1}
- 20\,\sqrt{\alpha^2 + 1}- 9\,\alpha\r)\,
\l(3\,\alpha\,k_0^2\,k+3\,i\,k_0^3\r)\biggr]\,
{\rm e}^{i\,\alpha\,k/k_0}\nn\\
& &+\, \alpha^2\,\sqrt{\alpha^2 + 1}\,\l(4\,3^{1/4}\,\pi\,k^2 + 9\,k_0\,k\r)\,
\l[\l(9 - 36\,\alpha^2\r)\,k_0 - i\,\sqrt{3}\,\alpha\,\l(\alpha^2 + 1\r)\,k\r]
{\rm e}^{i\,\alpha\,k/(\sqrt{3}\,k_0)}\nn\\
& &+\, \l(\alpha^2 + 1\r)\,\tan^{-1}(\alpha)\,
\biggl[3^{5/4}\,i\,\alpha^2\,\l(\alpha^2 + 1\r)\,
\l(-9\,\alpha^3 + 32\,\sqrt{\alpha^2 + 1} + 27\,\alpha\r)\,k_0\,k^2\nn\\ 
& &+\, 3^{9/4}\,\l(9\,\alpha^5- 18\,\alpha^3 + 16\,\sqrt{\alpha^2 + 1}\,\alpha^2 
- 80\,\sqrt{\alpha^2 + 1} - 27\,\alpha\r)\, 
\l(\alpha\,k_0^2\,k + i\,k_0^3\r)\biggr]\,
{\rm e}^{i\,\alpha\,k/k_0}\nn\\
& &+\, 4\,\alpha^2\,\l(\alpha^2 + 1\r)^{3/2}\,\tan^{-1}(\alpha)\,
\l(4\,3^{1/4}\,\pi\,k + 9\,k_0\r)\nn\\
& &\times\,\l[-9\,\l(\alpha^2 - 1\r)\,k_0\,k -\sqrt{3}\,i\,\alpha\, 
\l(\alpha^2 + 1\r)\,k^2\r]\,
{\rm e}^{i\,\alpha\,k/(\sqrt{3}\,k_0)}\Biggr\},
\end{eqnarray}
\begin{eqnarray}
\cE_k 
&=& \frac{{\rm e}^{-2\,\sqrt{5}\,\tan^{-1}(\alpha)}}{864\;3^{3/4}\,
\sqrt{10}\,\alpha^7\,\l(\alpha + \sqrt{5}\r)\,a_0\,k_0^2\,\Mp\,k^{3/2}}\,
\Biggl\{-16\;3^{1/4}\,\alpha^2\,\sqrt{\alpha^2 + 1}\,k^2\;
{\rm e}^{i\,\alpha\,k/(\sqrt{3}\,k_0)}\nn\\
& &\times\,\biggl[\sqrt{3}\,\alpha\left(\alpha^2 + 1\r)\,
\l(4\,\alpha^3 + 5\,\sqrt{5}\,\alpha^2 + 8\,\alpha + 3\,\sqrt{5}\r)\,k 
+ 9\,i\,\l(\sqrt{5}\,\alpha^2 + 8\,\alpha + 3\,\sqrt{5}\right)\,k_0\biggr]\nn\\
& &\times\,{\rm Ei}\l[i\,\l(3 - \sqrt{3}\r)\,\alpha\,k/(3\,k_0)\r]\nn\\
& &+\,\Biggl(-i\,\alpha^2\,\l(\alpha^2 + 1\r)\,
\biggl[-9\,\alpha^6 + 9\,\sqrt{5}\,\alpha^5 + 162\,\alpha^4 
+ \left(160\,\sqrt{5\,(\alpha^2 + 1)} + 171\right)\alpha^2 \nn\\
& &+\, \l(256\,\sqrt{\alpha^2 + 1} + 81\,\sqrt{5}\r)\,\alpha 
+ 96\,\sqrt{5\,(\alpha^2 + 1)} + 2\,\l(64\,\sqrt{\alpha^2 + 1} 
+ 45\,\sqrt{5}\r)\,\alpha^3\biggr]\,k^2 \nn\\
& &+\, \biggl[-9\,\alpha^8 + 9\,\sqrt{5}\,\alpha^7 + 153\,\alpha^6 
+ 19\,\l(16\,\sqrt{5\,(\alpha^2 + 1)} + 9\r)\,\alpha^2\nn\\ 
& &+\, \l(640\,\sqrt{\alpha^2 + 1} + 81\,\sqrt{5}\r)\,\alpha 
+ 240\,\sqrt{5\,(\alpha^2 + 1)} 
+ \l(128\,\sqrt{\alpha^2 + 1} + 99\,\sqrt{5}\r)\,\alpha^5\nn\\ 
& &+\, \left(160\,\sqrt{5\,(\alpha^2 + 1)} + 333\right)\alpha^4
+ 3\,\l(128\,\sqrt{\alpha^2 + 1} + 57\,\sqrt{5}\r)\,\alpha^3\biggr]\,
\l(3\,\alpha\,k_0\,k+3\,i\,k_0^2\r)\Biggr)\nn\\
& &\times\,3^{5/4}\,k_0\,{\rm e}^{i\,\alpha\,k/k_0}\nn\\
& &+\, 4\,i\,\alpha^2\,\sqrt{\alpha^2 + 1}\, 
\l(4\,3^{1/4}\,\pi\,k^2 + 9\,k_0\,k\r)\,
\biggl[9\,i\,\l(\sqrt{5}\,\alpha^2 + 8\,\alpha + 3\,\sqrt{5}\r)\,k_0\nn\\
& &+\,\sqrt{3}\,\alpha\,\l(\alpha^2 + 1\r)\,
\l(4\,\alpha^3 + 5\,\sqrt{5}\,\alpha^2 + 8\,\alpha + 3\,\sqrt{5}\r)\,k\biggr]\,
{\rm e}^{i\,\alpha\,k/(\sqrt{3}\,k_0)}\Biggr\},\nn\\
\end{eqnarray}
\begin{eqnarray}
\cF_k 
&=& \frac{{\rm e}^{2\,\sqrt{5}\,\tan^{-1}(\alpha)}}{864\;3^{3/4}\,\sqrt{10}\,
\alpha^7\,\l(\alpha + \sqrt{5}\r)\,a_0\,k_0^2\,\Mpl\,k^{3/2}}\,
\Biggl\{16\,3^{1/4}\,\alpha^2\,\sqrt{\alpha^2 + 1}\,k^2\nn\\
& &\times\,\biggl[\sqrt{3}\,\alpha\,\l(\alpha^2 + 1\r)\,
\l(4\,\alpha^3 + 3\,\sqrt{5}\,\alpha^2 - 2\,\alpha + 3\,\sqrt{5}\right)k 
- 9\,i\left(\sqrt{5}\,\alpha^2 + 2\,\alpha - 3\,\sqrt{5}\r)\,k_0\biggr]\nn\\
& &\times\,{\rm Ei}\l[i\,\l(3 - \sqrt{3}\r)\,\alpha\,k/(3\,k_0)\r]\,
{\rm e}^{i\,\alpha\,k/(\sqrt{3}\,k_0)}\nn\\
& &+\,\Biggl(i\,\alpha^2\left(\alpha^2 + 1\right)\biggl[-9\,\alpha^6 
- 27\,\sqrt{5}\,\alpha^5 - 18\,\alpha^4 
+ \l(96\,\sqrt{5\,(\alpha^2 + 1)} -9\r)\,\alpha^2\nn\\
& &+\, \l(81\,\sqrt{5} - 64\,\sqrt{\alpha^2 + 1}\r)\,\alpha 
+ 96\,\sqrt{5\,(\alpha^2 + 1)} 
+ 2\,\l(64\,\sqrt{\alpha^2 + 1} + 27\,\sqrt{5}\r)\,\alpha^3\biggr]k^2\nn\\
& &-\,\biggl[-9\,\alpha^8 - 27\,\sqrt{5}\,\alpha^7 - 27\,\alpha^6 
+ 9\,\l(16\,\sqrt{5\,(\alpha^2 + 1)} -1\r)\,\alpha^2\nn\\
& &+\, \l(81\,\sqrt{5} - 160\,\sqrt{\alpha^2 + 1}\r)\,\alpha
+ 240\,\sqrt{5\,(\alpha^2 + 1)} + \l(128\,\sqrt{\alpha^2 + 1} 
+ 27\,\sqrt{5}\r)\,\alpha^5\nn\\ 
& &+\, 3\,\left(32\,\sqrt{5\,(\alpha^2 + 1)}- 9\r)\,\alpha^4
+ \l(64\,\sqrt{\alpha^2 + 1} + 135\,\sqrt{5}\r)\,\alpha^3\biggr]\,
\l(3\,\alpha\,k_0\,k+3\,i\,k_0^2\r)\Biggr)\nn\\
& &\times\, 3^{5/4}\,k_0\,\,{\rm e}^{i\,\alpha\,k/k_0}\nn\\
& &-\, 4\,\alpha^2\,\sqrt{\alpha^2 + 1}\, 
\l(4\,3^{1/4}\,\pi\,k^2 + 9\,k_0\,k\r)\,
\biggl[9\,\l(\sqrt{5}\,\alpha^2 + 2\,\alpha - 3\,\sqrt{5}\r)\,k_0\nn\\ 
& &+\, \sqrt{3}\,i\,\alpha\,\l(\alpha^2 + 1\r)\,
\l(4\,\alpha^3 + 3\,\sqrt{5}\,\alpha^2 - 2\,\alpha + 3\,\sqrt{5}\r)\,k\biggr]\,
{\rm e}^{i\,\alpha\,k/(\sqrt{3}\,k_0)}\Biggr\}.
\end{eqnarray}
\bibliographystyle{JHEP}
\bibliography{sps-mb-december-2017}
\end{document}